%% file: icedd_arxiv.tex
\documentclass[a4paper, conference]{IEEEtran}
\IEEEoverridecommandlockouts
\usepackage[nocompress]{cite}
\usepackage{amsmath,amssymb,amsfonts}
\usepackage{algorithmic}
\usepackage{graphicx}
\usepackage{textcomp}
\usepackage{xcolor}
\usepackage{amssymb}  
\usepackage{algorithm}
\usepackage{cuted}
\usepackage{booktabs}
\usepackage[normalem]{ulem}

\begin{document}
    \input{sections/header}
    \input{sections/abstract}

    \input{sections/introduction}
    \input{sections/system_model}
    \input{sections/proposed}
    \input{sections/Tracking}
    \input{sections/Analysis}

    \input{sections/results}
    \newpage
    \input{sections/conclusion}

    \input{sections/appendix}
    \input{sections/appendix2}
    \vspace{-0.5cm}
    \input{sections/appendix_B}

    \vspace{-0.5cm}
    \bibliography{support/main}
\end{document}

%% file: sections/header.tex
\title{Iterative Channel Estimation, Detection and Decoding for Multi-Antenna Systems with RIS }

\author{Roberto C. G. Porto and Rodrigo C. de Lamare \vspace{-1.5em}

\thanks{The authors are with the Centre for Telecommunications Studies, Department of Electrical Engineering, Pontifical Catholic University of Rio de Janeiro. Emails: camara@ime.eb.br, delamare@puc-rio.br}}

\maketitle

%% file: sections/abstract.tex
\begin{abstract}

 {This work proposes an iterative channel estimation, detection and decoding (ICEDD)  scheme for the uplink of multi-user multi-antenna systems assisted by multiple reconfigurable intelligent surfaces (RIS)}. A novel iterative code-aided channel estimation (ICCE) technique is developed that uses low-density parity-check (LDPC) codes and iterative processing to enhance estimation accuracy while reducing pilot overhead. The core idea is to exploit encoded pilots (EP), enabling the use of both pilot and parity bits to iteratively refine channel estimates. To further improve performance, an iterative channel tracking (ICT) method is proposed that takes advantage of the temporal correlation of the channel. An analytical evaluation of the proposed estimator is provided in terms of normalized mean-squared error (NMSE), along with a study of its computational complexity and the impact of the code rate. Numerical results validate the performance of the proposed scheme in a sub-6 GHz multi-RIS scenario with non-sparse propagation, under both LOS and NLOS conditions, and different RIS architectures. 
 
\end{abstract}

\begin{IEEEkeywords}

Reconfigurable intelligent surfaces, channel estimation, multiple-antenna systems, iterative detection and decoding, channel tracking.
\end{IEEEkeywords}

%% file: sections/introduction.tex
\vspace{-1em}
\section{Introduction}

Reconfigurable Intelligent Surfaces (RIS) have attracted considerable attention as a transformative technology for sixth-generation (6G) wireless communication networks. In order to further enhance the performance of RIS, several extensions of the original passive RIS architecture (P-RIS) \cite{9140329} have been proposed, including Active RIS \cite{9998527}, Beyond Diagonal (BD) RIS \cite{10308579}, Stacked Intelligent Metasurfaces (SIM) \cite{10158690}, and Reconfigurable Holographic Surfaces (RHS) \cite{9696209}.

These advanced RIS architectures have demonstrated significant gains in communication performance. However, a major challenge persists across all these approaches: channel estimation \cite{10818440}. The large number of channel coefficients inherent in RIS-assisted systems imposes strict requirements on pilot signaling, especially as the number of users and RIS elements increases. In such cases, obtaining accurate channel estimates may become prohibitive, thereby limiting the scalability of these systems. This challenge is further intensified in multi-RIS scenarios \cite{10439018,10558715,10623725,10497119}, where the number of effective channels grows even larger, making reliable estimation with a limited pilot budget increasingly challenging.

\vspace{-0.10cm}
\subsection{Prior and Related Work}

Several studies have investigated channel estimation techniques for RIS-assisted systems. In \cite{10025776}, a deep learning-based approach was proposed and evaluated using a compressed sensing method. Similarly, \cite{9127834} studied a deep neural network to improve compressive channel estimation by applying a complex-valued denoising convolutional neural network in millimeter-wave systems. Meanwhile, \cite{10614235} employed Bayesian learning for channel estimation in massive multiple-input multiple-output (MIMO) systems. \textcolor{black}{A key insight in most channel estimation studies is that the RIS reflects signals from all users to the BS through the same propagation paths, which results in correlation among the user–RIS–BS reflected channels that can be exploited to reduce channel estimation overhead.} In \cite{10818440}, the authors proposed a two-timescale channel estimation strategy using deterministic modeling and \textcolor{black}{a maximum likelihood estimator to reduce pilot overhead without hardware complexity, whereas} \cite{9130088} introduced a three-phase framework that exploits channel correlation by disabling the RIS during the first phase. In contrast, \cite{9839429} introduces an ``always-ON" protocol, where the RIS remain active throughout the channel estimation process. This not only eliminates the need for on-off amplitude control but also offers a more realistic implementation. For dealing with multi-RIS MIMO systems, \cite{10767769} proposed a semi-blind PARAFAC-based approach for joint channel estimation and symbol detection using alternating least squares (LS). 

Recent efforts such as \cite{10484981} have considered pilot-aided estimation schemes with time-varying channels in RIS-assisted cell-free systems. Complementary approaches have focused on improving estimation accuracy in dynamic environments. In \cite{10250189}, a Kalman filter-based scheme was proposed to track time-varying RIS channels under user mobility and pilot contamination, which was further extended to handle hardware impairments using an extended Kalman filter. In \cite{9854102}, a two-stage transmission protocol combining Kalman filtering and deep learning was introduced to enable low-overhead channel tracking and prediction. Furthermore, \cite{10818463} explored deep learning-based channel prediction using active RIS elements, emphasizing the impact of element arrangement on estimation accuracy. Nonetheless, these approaches rely on uncoded systems, where channel estimation is performed solely based on pilot symbols, without leveraging the parity bits available in encoded data to enhance estimation performance.

{
To address this limitation, a few works have explored code-aided channel estimation techniques. In \cite{10319806}, a code-aided channel estimation method was proposed for LDPC-coded small-scale point-to-point MIMO systems under block-fading conditions. This method exploits the sparsity of the low-density parity-check (LDPC) parity-check matrix to approximate the likelihood of valid codewords and refine the channel coefficients through coordinate ascent optimization over the in-phase and quadrature components that are used to compute the log-likelihood ratios (LLRs). Although accurate, such approaches are computationally impractical for large-scale systems such as RIS-assisted or multiuser MIMO scenarios. In a related work, an iterative channel estimation and LDPC decoding scheme with encoded pilots was proposed in \cite{4357052} for SISO OFDM systems using BPSK modulation. However, this method does not exploit temporal channel correlation and relies solely on decoder feedback for refinement. Although encoded pilots (EP) are not strictly required for joint channel estimation and decoding, their known positions and values can be exploited by the decoder as reliable bits, allowing faster convergence or lower bit error rates in low-SNR regions.}

\subsection{Contributions}

{
Motivated by these observations and by our previous work in which an iterative detection and decoding (IDD) scheme was developed for RIS-aided systems  \cite{10747209}, this work extends that concept to jointly perform channel estimation. The earlier scheme combined LDPC codes with iterative refinement of the LLRs exchanged between detector and decoder. In this work, the proposed framework integrates channel estimation into the iterative process, jointly performing detection, decoding, and estimation by exploiting both encoded pilots and parity bits, thereby improving convergence speed and robustness to channel variations. In addition to decoder feedback, the receiver incorporates extrinsic information from previous estimates, enabling the channel to be refined iteratively. The use of encoded pilots is particularly beneficial when only a limited number of pilot symbols are available to obtain a coarse estimate of the user–RIS–BS reflected channels, which are subsequently refined through code-aided processing. Furthermore, the framework exploits temporal correlation between consecutive packets to further reduce the number of required coded pilot symbols for accurate channel estimation.}
{A preliminary version of this work was presented in \cite{spawc}. The current work significantly extends the previous one by incorporating a more comprehensive channel tracking approach, an improved decoding strategy, and a detailed performance evaluation under various mobility and channel conditions. Specifically, our contributions are summarized as follows:}


\begin{itemize}
    
    \item {We develop an iterative channel estimation, detection and decoding (ICEDD) scheme along with an efficient Iterative Code-Aided Channel Estimation (ICCE) algorithm for multi-RIS-assisted systems. The proposed framework exploits both line-of-sight (LOS) and non-line-of-sight (NLOS) components, while reducing pilot overhead by leveraging the coding structure. The proposed scheme and algorithm are compatible with various RIS architectures, enabling flexible deployments.}
    
    \item An iterative channel tracking (ICT) method with two-time-slot transmission protocol is introduced to exploit the temporal correlation of time-varying channels, further improving estimation accuracy and efficiency.
    
    \item {We provide an analytical study of the ICCE algorithm, including its complexity analysis and a study of its convergence behavior. The analysis is validated through extensive numerical results under various LOS and NLOS conditions, as well as different RIS configurations.}
\end{itemize}

The remainder of this paper is organized as follows. Section II presents the system model. Section III describes the ICCE technique. In Section IV, we extend the estimation scheme to ICT in temporally correlated channels. Section V analyzes the convergence and computational complexity of the proposed algorithms. Section VI discusses the simulation results and Section VII concludes the paper.

\textit{Notation:} Bold capital letters represent matrices, while bold lowercase letters denote vectors. The symbol $\mathbf{I_n}$ refers to an identity matrix $n \times n$, and $\text{diag}(\mathbf{A})$ is a diagonal matrix with the diagonal elements of $\mathbf{A}$. The sets of complex and real numbers are denoted by $\mathbb{C}$ and $\mathbb{R}$, respectively. [·]$^{-1}$, [·]$^{T}$, and [·]$^H$ denote the inverse, transpose, and conjugate transpose, respectively. The Kronecker product is represented by $\otimes$.


\begin{figure*}
    \centerline{\includegraphics[width=1.0\textwidth]{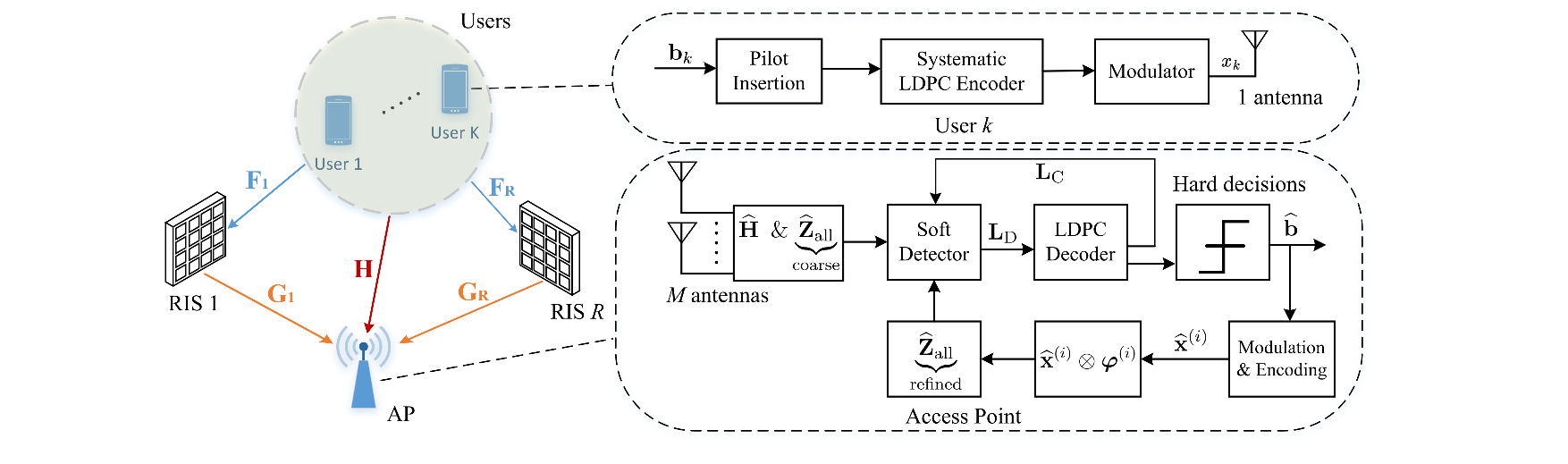}}
    \vspace{-0.75em}
    \caption{System model of an ICEDD multiuser multiple-antenna system.}
    \label{fig:blockdiagram}
    \vspace{-1.0em}
\end{figure*}


%% file: sections/system_model.tex
\section{System Model}

A single-cell multiuser {uplink} system featuring a multiple-antenna access point (AP) assisted by $R$ reflective intelligent surfaces (RIS) {arranged in a Uniform Planar Array (UPA)} is considered, as illustrated in Fig. \ref{fig:blockdiagram}. In this configuration, the AP is equipped with $M$ {UPA} antennas \cite{mmimo,wence} that support $K$ users, each equipped with a single antenna. 

For each user, the information bits and pilot bits are jointly encoded using individual systematic LDPC channel encoders \cite{memd}, and the resulting codewords are subsequently modulated into $x_k$ by user $k$, using a finite modulation scheme. The encoded packet has a fixed length of $N_\text{block}$. {We assume perfect symbol-level synchronization among all users} and, for clarity, refer to the pilot and parity components prior to modulation as pilot bits and parity bits, and to their corresponding modulated values as pilot symbols and parity symbols, respectively. The transmit symbols $x_k$ are zero-mean and have equal average energy, satisfying $\mathbb{E}[|x_k|^2] = \sigma_x^2$. These modulated symbols are transmitted over block-fading channels.

Each RIS has $L$ reflecting elements, and their reflection coefficients are modeled as a complex vector {$\boldsymbol{\varphi}_r \triangleq [e^{j\theta_1}, \dots, e^{j\theta_L}]^T$, where $\theta_l$ is the phase shift of the $l$-th element of the $r$-th RIS}. The corresponding diagonal reflection matrix at time instant $i$ is given by {$\mathbf{\Phi}_r^{(i)} \triangleq \text{diag}(\boldsymbol{\varphi}_r^{(i)})$}. The $M$-dimensional received signal {$\mathbf{y}^{(i)}$} at time instant $i$ is given by
\begin{equation}
    \mathbf{y}^{(i)} = (\mathbf{H} + 
    \sum_{r=1}^R\mathbf{G}_r\mathbf{\Phi}_r^{(i)}\mathbf{F}_r +
     \mathbf{E}^{(i)})
    \mathbf{x}^{(i)} + \mathbf{n}^{(i)},
    \label{eq01}
\end{equation}
where $\mathbf{H} \in \mathbb{C}^{M \times K}$, {$\mathbf{G}_r \in \mathbb{C}^{M \times L}$, and $\mathbf{F}_r \in \mathbb{C}^{L \times K}$} represent the communication links from the AP to the users, from the AP to {RIS-$r$}, and from {RIS-$r$} to the users, respectively. The vector {$\mathbf{x}^{(i)} \triangleq [x_1^{(i)}, \dots, x_K^{(i)}]^T$} represents the coded symbols transmitted by the users at time instant $i$ and $\mathbf{n}^{(i)} \sim \mathcal{CN}(\mathbf{0}_M, \sigma_n^2\mathbf{I}_M)$ represents the noise. 

The signal component resulting from reflections between different RIS at time instant $i$ is denoted by $\mathbf{E}^{(i)} \in \mathbb{C}^{M \times K}$. Due to the multiplicative fading effect \cite{9998527}, this element has a negligible effect on the received signal and is ignored in the remainder of this work. Note that RIS can also be deployed to reduce the effect of $\mathbf{E}^{(i)}$.

The contributions of different RIS to the users can be grouped in a concise expression by
\begin{align}
\mathbf{y}^{(i)} =&  [\mathbf{G}_1 \dots \mathbf{G}_R]
\begin{bmatrix}
\boldsymbol{\Phi}_1^{(i)} & \boldsymbol{0} & \boldsymbol{0} \\
\boldsymbol{0} & \ddots           & \boldsymbol{0} \\
\boldsymbol{0} & \boldsymbol{0}   & \boldsymbol{\Phi}_R^{(i)} \\
\end{bmatrix} 
\begin{bmatrix}
\mathbf{F}_1 \\
\vdots \\
\mathbf{F}_R
\end{bmatrix}\mathbf{x}^{(i)}
\\
&+ \mathbf{H}\mathbf{x}^{(i)} + \mathbf{n}^{(i)}. \nonumber 
\end{align}
Grouping the matrices of the communication links leads to
\begin{equation}
    \mathbf{y}^{(i)} = (\mathbf{H} + \mathbf{G}_e\boldsymbol{\Phi}_e^{(i)}\mathbf{F}_e)\mathbf{x}^{(i)} + \mathbf{n}^{(i)} = \mathbf{\bar{H}}_e\mathbf{x}^{(i)}  + \mathbf{n}^{(i)}, 
    \label{eq03}
\end{equation}
where $\color{black}\mathbf{\bar{H}}_e$ represents the equivalent channel between the AP and the users.

Eq. (\ref{eq03}) is similar to that used for the representation of a single RIS-assisted MIMO system, which differs only in how the matrices are grouped. Therefore, the same techniques can be used here. In addition to the grouped RIS phase-shift matrix, $\boldsymbol{\Phi}_e^{(i)}$ remains a diagonal-only matrix. Therefore, the received signal in (\ref{eq03}) can also be written in terms of $\color{black}\boldsymbol{\varphi}^{(i)}_e$ as given by 
\begin{equation}
     \mathbf{y}^{(i)} = \mathbf{H}\mathbf{x}^{(i)} + \sum_{k=1}^K\mathbf{Z}_k\boldsymbol{\varphi}_e^{(i)}x_k^{(i)} + 
     \mathbf{n}^{(i)},
    \label{eq:eq04}
\end{equation}
where $\mathbf{Z}_k = \mathbf{G}_e \text{diag}(\mathbf{f}_{e,k}) \in \mathbb{C}^{M\times L_e}$, with  $\mathbf{f}_{e,k}$  denoting the 
$k$th column of the matrix $\mathbf{F}_e$, and $L_e =LR$ represents the total number of RIS elements in the system.

An estimate $\hat{x}_k$ of the transmitted symbol is obtained by applying a linear receive filter {$\mathbf{w_k}$ to the received signal, without performing soft interference cancellation (SIC), as}
\begin{equation}
    \hat{x}_k^{(i)} = (\mathbf{w}_k^{(i)})^H\mathbf{y}^{(i)} = \left(\frac{\sigma^2_n}{\sigma^2_x}\mathbf{I_{n_r}} + \mathbf{\bar{H}}_e \mathbf{\bar{H}}_e^{\rm H} \right)^{-1}\mathbf{\bar{h}}_{e,k}\mathbf{y}^{(i)},
    \label{detection_estimate_1}
\end{equation}
{where  \( \mathbf{\bar{h}}_{e,k} \in \mathbb{C}^M \) and \( \mathbf{\bar{H}}_e \triangleq [\mathbf{\bar{h}}_{e,1}, \dots, \mathbf{\bar{h}}_{e,K}]^H \in \mathbb{C}^{M \times K} \).}
\subsection{Enhancing Detection Through SIC}
\label{subsec:SIC}
In the  {SIC} detector \cite{spa,mfsic,mbdf,rsbd,rsthp}, the received vector $\mathbf{y}^{(i)}$ is processed by demapping, where a LLR is calculated for each bit in the transmit vector $\mathbf{x}^{(i)}$. For simplicity, we omit the $(i)$ notation in this section. \textcolor{black}{Therefore, the extrinsic LLR value $L_D$ associated with the $\nu$-th bit $b_\nu$ in the binary code sequence is computed as follows}
\begin{equation}
    L_{D}(b_{\nu}) = \log \frac{\sum\nolimits_{\mathbf{x} \in \mathcal{X}_{\nu}^{+1}} P(\mathbf{y} \vert \mathbf{x}, \mathbf{\bar{H}}_e) P(\mathbf{x})} {\sum\nolimits_{\mathbf{x} \in \mathcal{X}_{\nu}^{-1}} P(\mathbf{y} \vert \mathbf{x}, \mathbf{\bar{H}}_e) P(\mathbf{x})} - L_{C}(b_{\nu}).
    \label{eq:ldlc}
\end{equation}
Inspired by prior work on IDD schemes \cite{spa,mfsic,mbdf,8240730,dfcc,bfidd,did,rrber,listmtc,msgamp,decidd,iddllr,iddocl}, the soft estimate of the $k$-th transmitted symbol is firstly calculated based on the $\mathbf{L}_C$ (extrinsic LLR) provided by the channel decoder from a previous stage:
\begin{equation*} 
\tilde {x}_{k}=\sum _{x\in \mathcal {A}}x\text {Pr}(x_{k}=x)=\sum _{x\in \mathcal {A}}x\left ({\prod _{l=1}^{M_{c}}\left [{1+\text {exp}(-x^{l}L_{c}^{l})}\right]^{-1}}\right), 
\end{equation*}
where $\mathcal {A}$ is the complex constellation set with $2^{M_c}$ possible points. The symbol $x^l$ corresponds to the value $(+1, -1)$ of the $l$th bit of the symbol $x$. 

A symbol estimate uses SIC, where the value of $\boldsymbol{\varphi}_e \in \mathbb{C}^{L_e}$ is fixed and $\mathbf{w_k} \in \mathbb{C}^M$ is chosen to minimize the mean square error (MSE) between the transmitted symbol $x_k$ and the filter output
{\begin{equation} 
    \mathbf {w}_{k}=\arg \min _{ \tilde{\mathbf{w}}_{k}} E\left [{\left \vert{ x_{k}-\tilde{\mathbf{w}}_{k}^{H}\mathbf {y}_k}\right \vert ^{2}}\right]. 
\end{equation}}
It can be shown that the solution is given by
\begin{equation}
    \mathbf{w}_k = \left(\frac{\sigma^2_n}{\sigma^2_x}\mathbf{I_{n_r}} + \mathbf{\bar{H}}_e \mathbf{\Delta}_k\mathbf{\bar{H}}_e^{\rm H} \right)^{-1}\mathbf{\bar{h}}_{e,k},
    \label{eq:w}
\end{equation}
where \( \mathbf{\bar{H}}_e \in \mathbb{C}^{M \times K} \), \( \mathbf{\bar{h}}_{e,k} \in \mathbb{C}^M \), and the covariance matrix \( \mathbf{\Delta}_k \in \mathbb{C}^{K \times K} \) is

\begin{equation}
    \mathbf{\Delta}_k = \text{diag}\left[\frac{\sigma^2_{x_{1}}}{\sigma^2_x}\dots \frac{\sigma^2_{x_{k-1}}}{\sigma^2_x}, 1, \frac{\sigma^2_{x_{k+1}}}{\sigma^2_x},\dots,\frac{\sigma^2_{x^2_{K}}}{\sigma^2_x}  \right],
\end{equation}
where $\sigma^2_{x_{i}}$ is the variance of the $k$th user that is computed by
{\begin{equation}
\sigma_{x_{k}}^{2}=\sum\limits_{x\in {\cal A}}\vert x-\bar{x} _{i}\vert ^{2}P(x_{k}=x). 
\end{equation}}

\subsection{Design of Reflection Parameters}

The computation of reflection parameters builds upon the work in \cite{10747209}, which focuses on their design based on the minimum mean square error (MMSE) criterion. While that work considers a single RIS, we generalize the approach to a scenario with multiple RIS. This MMSE design refines LLRs within an IDD system using MMSE receive filters that are used to facilitate SIC at the receiver. The reflection parameters computed by the MMSE criterion are given by
\vspace{-0.5em}
\begin{equation} 
\boldsymbol{\varphi}_o = \mathbf{B}^{-1}\boldsymbol{\Psi},
\label{eq05}
\end{equation}
\vspace{-0.5em}
where
\begin{equation}
    \mathbf{B} = \sum^{K}_{k=1}(\mathbf{WZ}_k)^H(\mathbf{WZ}_k),
\end{equation}
\vspace{-0.5em}
\begin{equation}
    \boldsymbol{\Psi} = \sum_{k=1}^{K} (\mathbf{WZ}_k)^H (\mathbf{e}_k - \mathbf{W} \mathbf{\bar{h}}_k),
\end{equation}
where $\mathbf {W} \triangleq [\mathbf{w_1}, \dots, \mathbf{w_K}]^H \in \mathbb{C}^{K\times M}$ represents the reception filter matrix, $\mathbf{e}_k\in \mathbb{C}^{K}$ is a column vector with zeros, except for the one in the $k$-th element, \(\mathbf{B} \in \mathbb{C}^{{L_e}\times {L_e}}\), and \(\boldsymbol{\Psi} \in \mathbb{C}^{L_e}\). This method involves truncating the reflection parameters to satisfy the RIS constraint  $|[\boldsymbol{\varphi}]_n|=1$, for $\forall n$, which leads to the following truncation
\begin{equation}
    [\boldsymbol{\varphi}_{o,\mathrm{trunc}}]_i = \frac{[{\varphi}_o]_i}{|[{\varphi}_o]_i|}  \Leftrightarrow  \boldsymbol{\varphi}_{o,\mathrm{trunc}}= e^{j\measuredangle (\boldsymbol{\varphi_0})}.
\end{equation}

Although this design does not meet the MMSE requirement in a strict sense, it represents the closest feasible solution under the unit-modulus constraint. This phase-projection approach preserves most of the performance of the unconstrained MMSE solution while maintaining low implementation cost.

In practice, the reflection parameters are computed using the estimated channel coefficients ($\hat{\mathbf{H}}$, $\hat{\mathbf{G}}$, $\hat{\mathbf{F}}$) rather than the actual CSI. In this work, the value obtained from the estimated channels is denoted as $\boldsymbol{\varphi}_o$, since it is fully determined at the AP and is selected based on the channel estimates, even though it may not be optimal due to estimation and truncation effects.

\subsection{Time-varying Channel Model}
The flat-fading MIMO channel is modeled as a discrete-time first-order Gauss-Markov process, where the channel remains constant within each transmission block and evolves over time. This reflects scenarios where the block duration is shorter than or comparable to the channel coherence time~\cite{10484981}.

For simplicity and considering a worst-case scenario, we assume all users are equidistant from the AP and have similar mobility. Thus, a common temporal correlation coefficient \(\rho \in [0,1]\) is adopted. The evolution of a generic channel matrix—representing either a direct or reflected link—is given by
\begin{equation}
    \mathbf{H}_{\chi}^{(b)} = \rho\, \mathbf{H}_{\chi}^{(b-1)} + \sqrt{1 - \rho^2}\, \mathbf{\Gamma}_{\chi}^{(b)},
    \label{eq:markov_common}
\end{equation}
where \(\chi \in \{\text{direct}, \text{reflected}\}\) denotes the channel type, and \(\mathbf{\Gamma}_{\chi}^{(b)} \sim \mathcal{CN}(0, \mathbf{R}_{\chi})\), and \(\mathbf{R}_{\chi}\) is the covariance matrix of \( \mathbf{H}_{\chi}^{(b)}\).




%% file: sections/proposed.tex
\section{Iterative Code-Aided Channel Estimation}
\label{sec:proposed}
\begin{figure}
\centerline{\includegraphics[width=0.5\textwidth]{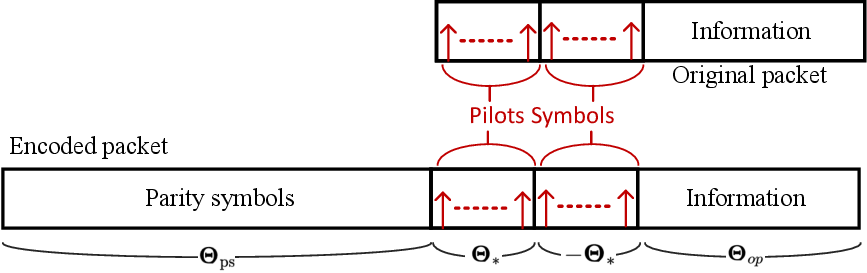}}
    \vspace{-0.725em}
    \caption{Systematically encoded packet and original post-modulation packet.}
    \label{fig:package}
    \vspace{-1em}
\end{figure}

In this section, we propose a novel channel estimation technique for multi-RIS-assisted {uplink} systems that leverages iterative detection and decoding with an encoded pilot  and LDPC decoding to reduce the number of pilots required to estimate the cascaded RIS coefficients. All the pilots symbols are encoded with the data bits using a systematic encoder; therefore, the pilots symbols remain unaltered and can be considered known by the receiver, as illustrated in Fig. \ref{fig:package}. 
The technique begins by estimating the direct channel and obtaining an initial coarse estimation of the reflected RIS channel. After the first iteration, the entire packet with estimated symbols is used as pilots to refine the estimation of the reflected channel. Since the receiver is prone to incorrect symbol estimates, this procedure is performed iteratively to refine the channel estimation in each iteration.

To introduce this approach, we first explain how the direct channel can be estimated, then describe the coarse estimation of the reflected channel, and eventually show how EP and the iterative processing enhance the overall estimation accuracy.

\subsection{Direct Channel Estimation}
\label{subsec:direct}
To estimate the direct channel, we used an always-on channel estimation approach without switching off the selected RIS elements \cite{9839429}. This protocol involves dividing the pilot sequence into two partitions of equal size, allowing us to represent the received signal for each partition as
\begin{equation}
    \mathbf{y}^{(p)} = (\mathbf{H} + \mathbf{G}_e\boldsymbol{\Phi}_e^{(p)}\mathbf{F}_e)\mathbf{x}^{(p)} + \mathbf{n}^{(p)},
    \label{eq:partA}
\end{equation}
where \textcolor{black}{$p$} belongs to the set of pilot symbols. We define the first and second partitions, respectively, as:  
\begin{equation}
    \mathcal{P}_1 = \left\{ p \mid t \leq p \leq t + \frac{N_p}{2}-1\right\}    
    \label{eq:partition01}
\end{equation}
\begin{equation}
    \mathcal{P}_2 = \left\{ p \mid t + \frac{N_p}{2} \leq p \leq t + N_p -1\right\}.    
\label{eq:partition02}
\end{equation}
where \(N_p\) denotes the total number of pilot symbols.

By selecting the pilots and reflection parameter values such that \textcolor{black}{$x^{(p)} = x^{\left(p + \frac{N_p}{2}\right)}$  and $\boldsymbol{\Phi}_e^{(p)} = -\boldsymbol{\Phi}_e^{\left(p + \frac{N_p}{2}\right)}$}, we can define the sum and subtraction of each received signal as

\begin{equation}
    \frac{\mathbf{y}^{(p)}+\mathbf{y}^{(p+\frac{Np}{2})}}{2} = \mathbf{H}\mathbf{x}^{(p)} + \mathbf{u}^{(p)}
    \label{eq:directchannel}
\end{equation}
and
\begin{equation}
    \frac{\mathbf{y}^{(p)}-\mathbf{y}^{(p+\frac{Np}{2})}}{2} = \textbf{G}_p\boldsymbol{\Phi}_e^{(p)}\textbf{F}_p\mathbf{x}^{(p)} + \mathbf{u}^{(p)},
    \label{eq:reflectedchannel}
\end{equation}
where $\mathbf{u} \sim \mathcal{CN}(\mathbf{0_M}, \frac{\sigma_n^2}{2} \mathbf{I_M})$.

From (\ref{eq:directchannel}), we can apply conventional channel estimation methods to estimate only the direct channel. In this work, we use the linear minimum mean square error (LMMSE) channel estimator given by 
\begin{equation}
    \hat{\mathbf{H}}_\text{LMMSE} = \mathbf{Y}_p\left(\mathbf{P}^H\mathbf{R}_\text{H}\mathbf{P}+\frac{\sigma_n^2}{2\sigma_x^2}\mathbf{I}\right)^{-1}\mathbf{P}^H\mathbf{R}_\text{H},
    \label{eq:lmmse}
\end{equation}
{
where \( \mathbf{R}_\text{H} := \mathbb{E}[\mathbf{H}\mathbf{H}^H] \in \mathbb{C}^{M\times M} \) denotes the channel covariance matrix.  
The pilot matrix and received signal matrix are, respectively, given by
\begin{equation}
    \mathbf{P} = [\mathbf{p}_{s,1}, \mathbf{p}_{s,2}, \dots, \mathbf{p}_{s,N_p/2}] \in \mathbb{C}^{K \times N_p/2},
\end{equation}
\begin{equation}
    \mathbf{Y}_p = \left[\frac{\mathbf{y}^{(p)}+\mathbf{y}^{(p+\frac{Np}{2})}}{2}, \dots, \frac{\mathbf{y}^{(p+\frac{Np}{2})}+\mathbf{y}^{(N_p)}}{2}\right] \in \mathbb{C}^{M \times N_p/2},
\end{equation}
where \( \mathbf{p}_{s,p} \) denotes the  $p $-th transmitted pilot symbol vector, whose elements belong to the constellation set \( \mathcal{A} \), and the columns of \( \mathbf{P} \) are orthogonal to each other. For this estimation procedure, both pilot partitions are utilized, requiring a total of \( N_p = 2K \) pilot symbols.
}


{\subsection{Cascaded RIS Channel Estimation}}
\label{subsec:cascate}
To estimate the coefficients of the reflected channel without interference from the direct link, we use (\ref{eq:reflectedchannel}). By rewriting the received signal in terms of $\boldsymbol{\varphi}_e^{(i)}$ using (\ref{eq:eq04}), the received signal from the RIS can be expressed as

\begin{equation}
    \mathbf{y}^{(p)}_\text{cascaded}=\frac{\mathbf{y}^{(p)}-\mathbf{y}^{(p+\frac{Np}{2})}}{2} = \sum_{k=1}^K\mathbf{Z}_k\boldsymbol{\varphi}_e^{(p)}x_k^{(p)} + \mathbf{u}^{(p)}.
    \label{eq:cascate_1}
\end{equation}

Eq. \eqref{eq:cascate_1} can be rearranged by concatenating the matrices $\mathbf{Z}_k$ and eliminating the summation, resulting in 

\begin{align}
    \label{eq:cascate_1_red}
    \mathbf{y}^{(p)}_\text{cascaded}&= [\mathbf{Z}_1, \dots, \mathbf{Z}_K](\mathbf{x}^{(p)}\otimes\boldsymbol{\varphi}_e^{(p)} ) + \mathbf{u}^{(p)}
    \\ 
    &= \mathbf{Z_\text{all}}\boldsymbol{\lambda}^{(p)} + \mathbf{u}^{(p)}, \nonumber 
\end{align}
where $\mathbf{Z_\text{all}} \in \mathbb{C}^{M \times KL_e}$ is the concatenated matrix of the cascaded channel for each user and \textcolor{black}{$\boldsymbol{\lambda}^{(p)}=(\mathbf{x}^{(p)}\otimes\boldsymbol{\varphi}_e^{(p)} )$} is a complex vector with \(KL_e\) elements.

Since the symbols of partition $\mathcal{P}_1$ are known by the receiver, we can assume that \textcolor{black}{$\boldsymbol{\lambda}^{(p)}$} is also fully known. After $N_p$ pilot transmission time slots, we can obtain the $M \times N_p$ overall measurement matrix $\mathbf{Y}_\text{cascaded}=[\mathbf{y}_\text{cascaded}^{(1)}, \dots,\mathbf{y}_\text{cascaded}^{(N_p/2)}]$ as 
\begin{equation}
    \mathbf{Y}_\text{cascaded}= 
    \mathbf{Z_\text{all}}\boldsymbol{\Lambda}_p + \mathbf{U},  
    \label{eq:cascade_2}
\end{equation}
where $\boldsymbol{\Lambda}_p = [\boldsymbol{\lambda}^{(1)}, \dots, \boldsymbol{\lambda}^{(N_p/2)}] \in \mathbb{C}^{KL_e \times N_p/2}$ and $\mathbf{U} = [\mathbf{u}^{(1)}, \dots, \mathbf{u}^{(N_p/2)}] \in \mathbb{C}^{M \times N_p/2}$.

This expression can be considered equivalent to a conventional MIMO channel estimation problem, which allows us to apply (\ref{eq:lmmse}) to estimate the coefficients. 

To mitigate multiuser interference, the matrix $\boldsymbol{\Lambda}_p$ should be orthogonal or semi-orthogonal, ensuring that the coefficients of $\mathbf{Z_\text{all}}$ can be estimated independently for each user. This matrix should be well conditioned to prevent noise amplification when inverted. To this end, we generate $\boldsymbol{\lambda}^{(i)}$ using Hadamard sequences for $\mathbf{x}^{(i)}$, while $\boldsymbol{\varphi}_e^{(i)}$ is derived from the DFT matrix \cite{9130088}. Note that to estimate all channel coefficients, $\boldsymbol{\Lambda}_p$ must have at least a rank of $KL_e$, which may be infeasible in some scenarios. Therefore, we propose to first obtain a coarse approximation of the cascaded RIS channel using only a small number of pilot symbols, specifically, an effective number of pilots \( (N_p/2) \ll KL_e \), and subsequently refine the channel estimates through an iterative estimation procedure.

{Using an LMMSE channel estimator in (\ref{eq:cascade_2}), the coarse channel estimate is given by
\begin{equation}
\label{eq:cascaded_coarse}
    \hat{\mathbf{Z}}_{\text{all,coarse}} = \mathbf{Y}_{\text{reflected}} 
        \left( \boldsymbol{\Lambda}_p^H \mathbf{R}_H \boldsymbol{\Lambda}_p + \frac{\sigma_n^2}{2 \sigma_x^2} \mathbf{I} \right)^{-1} 
        \boldsymbol{\Lambda}_p^H \mathbf{R}_H .
\end{equation}
}

\subsection{Iterative Estimation}
\label{sec:proposedInt}
In this step, our goal is to use the entire decoded symbol sequence as input, even if some symbols are incorrectly decoded. As shown in Fig. \ref{fig:blockdiagram}, the output of the IDD scheme provides the decoded bits. {By adopting the EP scheme \cite{4357052}, which combines pilot and information bits through a systematic encoder, the EPs can be exploited to assist the iterative channel estimation and decoding. }

For iterative channel estimation, the first step is to reapply coding and modulation to transform the decoded bits back into symbols. Then, we compute the Kronecker product \textcolor{black}{$(\hat{\mathbf{x}}^{(i)}\otimes\boldsymbol{\varphi}_e^{(i)} )$ to obtain   $\hat{\boldsymbol{\Lambda}}$, which enables the application  of the LMMSE channel estimator in  (\ref{eq:cascade_2})}. However, it is crucial for $\hat{\boldsymbol{\Lambda}}$ to remain semi-orthogonal and well-conditioned. Since the phase configurations that satisfy these conditions differ from those obtained using (\ref{eq05}), we adopt a suboptimal approach for the parity bits.

Assuming that the packet in Fig. \ref{fig:package} consists of $N_\mathrm{ps}$ parity symbols, $N_\mathrm{p}$ pilot symbols and $N_\mathrm{info}$ information symbols, and that $\boldsymbol{\Theta}_\mathrm{i} = [\boldsymbol{\varphi}_e^{(i)}, \boldsymbol{\varphi}_e^{(i+1)}, \dots]$ represents the concatenation of the reflection parameter vectors, we can express the reflection parameter vectors for each symbol as follows:
\begin{equation}
    \begin{matrix}  
    {\boldsymbol{\Theta}}_{\mathrm{ps}} =\left [{
\begin{array}{cccc} 
1 &1  &\cdots &1 \\ 
1 &\tau  &\cdots &\tau ^{N_\mathrm{ps}-1} \\ 
\vdots  &\vdots &\ddots &\vdots \\ 
1 &\tau ^{L_e-1}  &\cdots &\tau ^{(L_e-1)(N_\mathrm{ps}-1)} \end{array}}\right]\end{matrix},
\end{equation}
\begin{equation}
    \boldsymbol{\Theta}_\mathrm{p} = [\boldsymbol{\Theta}_\mathrm{*} -\boldsymbol{\Theta}_\mathrm{*}],
\end{equation}
\begin{equation}
    \begin{matrix}  
    {\boldsymbol{\Theta}}_{*} =\left [{
\begin{array}{cccc} 
1 &1  &\cdots &1 \\ 
1 &\varpi  &\cdots &\varpi ^{N_\mathrm{p}/2-1} \\ 
\vdots  &\vdots &\ddots &\vdots \\ 
1 &\varpi ^{L_e-1}  &\cdots &\varpi ^{(L_e-1)(N_\mathrm{p}/2-1)} \end{array}}\right]\end{matrix},
\end{equation}

\begin{equation}
    \boldsymbol{\Theta}_{\mathrm{o}} = [\boldsymbol{\varphi}_\mathrm{o} 
    \dots
    \boldsymbol{\varphi}_\mathrm{o}], \quad
    \boldsymbol{\varphi}_\mathrm{o} \text{ computed with (\ref{eq05})}
\end{equation}
where \textcolor{black}{$\tau = e^\frac{-2\pi j}{N_\mathrm{ps}}$ and $\varpi = e^\frac{-4\pi j}{N_\mathrm{p}}$.} If $L_e<\frac{N_\mathrm{p}}{2}$, to ensure pseudo-orthogonality between the RIS elements, we concatenate $L_e \times L_e$ DFT matrices such that the condition $qL_e \geq \frac{N_\mathrm{p}}{2}$ is met, where $q \in \mathbb{N}^*$ is the number of concatenated matrices. 
This ensures pseudo-orthogonality in $\boldsymbol{\Lambda}$ for both sequences of parity bits and pilots, while preserving the information symbols with the optimal reflection parameters. Note that since the same optimal phase configuration is applied to the information symbols, $\boldsymbol{\Theta}_{\mathrm{o}}$ is a low-rank matrix, which minimally contributes to the channel estimation.

{Two signal models are considered for estimating the cascaded channels. In the first signal model, only the pilot symbols are used to obtain a coarse estimate using (\ref{eq:cascaded_coarse}), which is less affected by noise since the equivalent noise power is reduced according to (\ref{eq:reflectedchannel}).  In contrast, the second model employs all transmitted symbols to obtain a refined estimate as}
\begin{equation}
    \hat{\mathbf{Z}}_{\text{all}} = \mathbf{Y}_{\text{casc}} 
        \left( \boldsymbol{\hat{\Lambda}}^H \mathbf{R}_H \boldsymbol{\hat{\Lambda}} + \frac{\sigma_n^2}{\sigma_x^2} \mathbf{I} \right)^{-1} 
        \boldsymbol{\hat{\Lambda}}^H \mathbf{R}_H .
        \label{eq:Z_ref}
\end{equation}

The initial symbol estimates may contain errors, which can propagate to the channel estimation. To mitigate this effect, the channel estimates are iteratively refined and the symbols are reprocessed until convergence is achieved or a predefined maximum number of iterations is reached. The pseudo-code of the ICCE algorithm is shown below.


\begin{algorithm}[H]
\caption{Encoded Pilot Packet Generation}
\label{alg:generation}
\begin{algorithmic}[1]
    \REQUIRE Input bitstream $\mathbf{b}_{k}$ for each user $k$
    \ENSURE Symbol stream for user $k$: $[x_k^{(1)}, \dots, x_k^{(N_{\text{ps}} + N_p + N_{\text{info}})}]$
    
    \FOR{$k = 1$ to $K$}
        \STATE Generate auxiliary pilot symbols for user $k$: 
        \textcolor{black}{$\mathbf{x}_{\text{aux}} = S_\mathcal{A} \cdot \text{Hadamard}(k)$}
        
        \IF{LOS system}
            \STATE Construct pilot vector: \textcolor{black}{$\mathbf{x}_p = [\mathbf{x}_{\text{aux}}; \mathbf{x}_{\text{aux}}]$}
        \ELSIF{NLOS system}
            \STATE Construct pilot vector: \textcolor{black}{$\mathbf{x}_p = [\mathbf{x}_{\text{aux}}]$}
        \ENDIF
        
        \STATE Generate pilot bits: \textcolor{black}{$\mathbf{b}_p = \text{demodulate}(\mathbf{x}_{p})$}
        
        \STATE Assemble pilot+data bitstream: $\mathbf{b}_{\text{packet},k} = [\mathbf{b}_p \: ; \: \mathbf{b}_{k}]$
        
        \STATE Encode: $\mathbf{c}_k = \text{encode}(\mathbf{b}_{\text{packet},k})$
        
        \STATE Modulate: $[x_k^{(1)}, \dots, x_k^{(N_{\text{ps}} + N_p + N_{\text{info}})}] = \text{modulate}(\mathbf{c}_k^\top)$
    \ENDFOR
\end{algorithmic}
\vspace{0.5em}
{\scriptsize 
\textbf{Observation:} $S_\mathcal{A}$ represents an arbitrary symbol from the constellation $\mathcal{A}$, and $\text{Hadamard}(k)$ refers to the $k$-th row of a Hadamard matrix. The resulting pilot vectors are mutually orthogonal across users, facilitating interference-free separation at the receiver.
}
\end{algorithm}

\vspace{-0.5cm}
\begin{algorithm}[H]
\caption{Channel Estimation and Refinement (LOS)}
\footnotesize
\label{alg:generation2}
\begin{algorithmic}[1]
    \REQUIRE $\boldsymbol{\Lambda}$, ${\boldsymbol{\Theta}}_{\mathrm{ps}}$, ${\boldsymbol{\Theta}}_{*}$, ${\boldsymbol{\Theta}}_{o}$
    \ENSURE Estimated channels: $\hat{\mathbf{H}}$, $\hat{\mathbf{Z}}_{\mathrm{all}}$
    
    \STATE \textbf{Receive Signals:} Obtain \textcolor{black}{$\mathbf{y}^{(i)}$ and $\mathbf{y}^{(i+\frac{N_p}{2})}$} based on partitions $\mathcal{P}_1$ and $\mathcal{P}_2$
    
    \STATE \textbf{Direct Component:}
    \textcolor{black}{
    \begin{itemize}        
        \item Compute: $\mathbf{y}^{(i)}_{\text{direct}} = \dfrac{\mathbf{y}^{(i)} + \mathbf{y}^{(j+\frac{N_p}{2})}}{2}$
        \textcolor{black}{
        \item Form matrix: $\mathbf{Y}_p = \left[ \mathbf{y}^{(i)}_{\text{direct}}, \dots, \mathbf{y}^{(i - 1 + \frac{N_p}{2})}_{\text{direct}} \right]$}
        \item Estimate direct channel via LMMSE:
        \[
        \hat{\mathbf{H}} = \mathbf{Y}_p 
        \left( \mathbf{P}^H \mathbf{R}_H \mathbf{P} + \frac{\sigma_n^2}{2 \sigma_x^2} \mathbf{I} \right)^{-1} 
        \mathbf{P}^H \mathbf{R}_H
        \]
    \end{itemize}}
    
    \STATE \textbf{Reflected Component:}
    \textcolor{black}{
    \begin{itemize}        
        \item Compute: $\mathbf{y}^{(i)}_{\text{reflected}} = \dfrac{\mathbf{y}^{(i)} - \mathbf{y}^{(i+\frac{N_p}{2})}}{2}$
        \item Form matrix: $\mathbf{Y}_{\text{reflected}} = \left[ \mathbf{y}^{(i)}_{\text{reflected}}, \dots, \mathbf{y}^{(i - 1 + \frac{N_p}{2})}_{\text{reflected}} \right]$
        \item Coarse channel estimation:
        \vspace{-0.2cm}
        \[
        \hat{\mathbf{Z}}_{\text{all,coarse}} = \mathbf{Y}_{\text{reflected}} 
        \left( \boldsymbol{\Lambda}_p^H \mathbf{R}_H \boldsymbol{\Lambda}_p + \frac{\sigma_n^2}{2 \sigma_x^2} \mathbf{I} \right)^{-1} 
        \boldsymbol{\Lambda}_p^H \mathbf{R}_H
        ,\]
        where $\boldsymbol{\Lambda}_p = [\boldsymbol{\lambda}_p^{(1)}, \dots, \boldsymbol{\lambda}_p^{(N_p/2)}]$ and $\boldsymbol{\lambda}_p^{(i)}=(\mathbf{x}_p^{(i)}\otimes\boldsymbol{\varphi}_*^{(i)} )$.
    \end{itemize}}
    
    \vspace{0.1cm}
    
    \FOR{$t = 1$ to Max Iterations or until $\text{NMSE} < \text{tol}$}
        \vspace{0.1cm}
        \STATE \textbf{IDD Scheme (SIC):} Apply the iterative detection and decoding as described in Sec.~\ref{subsec:SIC}
        \vspace{0.1cm}
        \STATE \textbf{Refinement:}
        \begin{itemize}
            \item Build:
            \vspace{-0.1cm}
            \begin{equation*}
                \boldsymbol{\hat{\Lambda}} = {\left[ \boldsymbol{\hat{x}}^{(1)} \otimes \boldsymbol{\varphi}_\text{ps}^{(1)}, \dots, \mathbf{\hat{x}}^{(N_\text{ps}+N_\text{p}+N_\text{info})} \otimes \boldsymbol{\varphi}_o^{(N_\text{ps}+N_\text{p}+N_\text{info})} \right]}
            \end{equation*}
            \vspace{-0.2cm}
            \item Refined channel estimation:
            \vspace{-0.1cm}
            \[
            \hat{\mathbf{Z}}_{\text{all}} = \mathbf{Y}_{\text{casc}} 
            \left( \boldsymbol{\hat{\Lambda}}^H \mathbf{R}_H \boldsymbol{\hat{\Lambda}} + \frac{\sigma_n^2}{\sigma_x^2} \mathbf{I} \right)^{-1} 
            \boldsymbol{\hat{\Lambda}}^H \mathbf{R}_H
            \]
        \end{itemize}
        \vspace{-0.2cm}
    \ENDFOR
\end{algorithmic}
\vspace{0.5em}
{\scriptsize \textbf{Observation:} In the NLOS case, the process starts directly from the reflected component, as \textcolor{black}{\(\mathbf{y}^{(i)}_{\text{direct}} = \mathbf{0}\)} and \textcolor{black}{\(\mathbf{x}_p = [\mathbf{x}_{\text{aux}}]\)}; that is, the pilot is not repeated.}
\end{algorithm}


%% file: sections/Tracking.tex
\begin{figure*}
    \centerline{\includegraphics[width=0.975\textwidth]{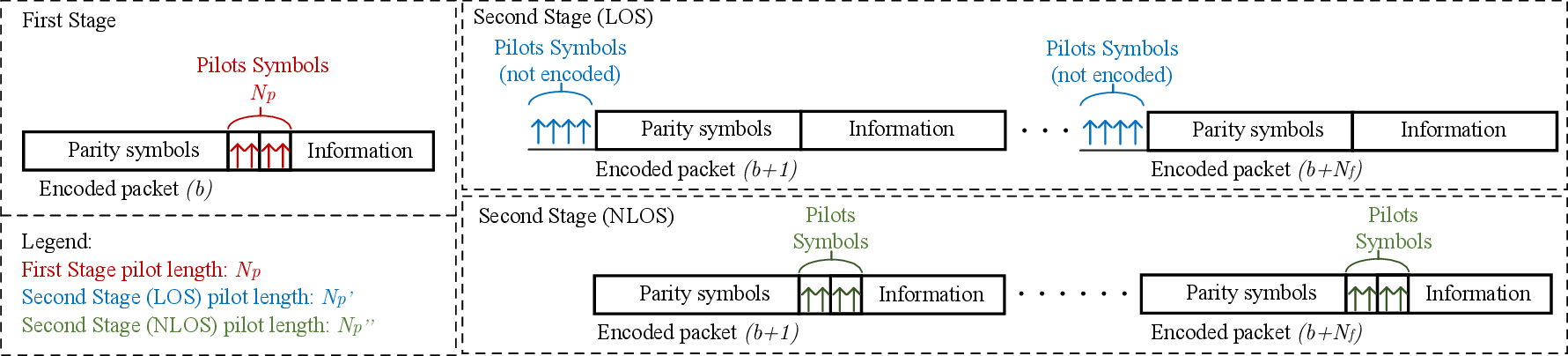}}
    \vspace{-0.75em}
    \caption{ICT Protocol for LOS and NLOS scenarios.}
    \label{fig:protocol}
    \vspace{-1.2em}
\end{figure*}

\newpage
\section{Iterative Channel Tracking for Temporal Correlated Channels}
\label{sec:temporal}

In this section, we propose exploiting the temporal correlation between adjacent transmitted blocks to improve the efficiency of the ICCE algorithm. Assuming that an enhanced channel estimate from the previous time instant, obtained through the ICCE algorithm, is available, the current reflected channel can be estimated as
\begin{equation}
    \hat{\mathbf{Z}}_\text{all}^{(b)} = \mathbf{Z}_\text{all}^{(b-1)} + \mathbf{E}^{(b-1)}_{\rm{est}}
\end{equation}
where $\mathbf{E}^{(b-1)}_{\rm{est}}$ denotes the estimation error associated with the previous block (indexed by $b{-}1$). Under the assumption that all channels are independent and identically distributed (\(\mathbf{R}_{\rm{refl}} = \sigma_z^2\mathbf{I}_M\)), and using the time-varying model in (\ref{eq:markov_common}), the normalized mean-square error (NMSE) associated with this estimate is given by
\begin{align}
    \varepsilon_l 
    &= \frac{ \mathbb{E} \left[ \left\| \mathbf{Z}_\text{all}^{(b)} - \hat{\mathbf{Z}}_\text{all}^{(b)} \right\|_F^2 \right] }
            { \mathbb{E} \left[ \left\| \mathbf{Z}_\text{all}^{(b)} \right\|_F^2 \right] }  \\
    &= \frac{ \mathbb{E} \left[ \left\| 
        \rho \mathbf{Z}_\text{all}^{(b-1)} 
        + \sqrt{1 - \rho^2}\, \mathbf{\Gamma}_{\text{refl}}^{(b)} 
        - \left( \mathbf{Z}_\text{all}^{(b-1)} + \mathbf{E}_\text{est}^{(b-1)} \right)
        \right\|_F^2 \right] }
        { \mathbb{E} \left[ \left\| \mathbf{Z}_\text{all}^{(b)} \right\|_F^2 \right] } \nonumber \\ \nonumber
    &\approx 2(1 - \rho) + \bar{\sigma}_\text{est}^2, 
    \quad \text{where} \quad 
    \bar{\sigma}_\text{est}^2 = \frac{ \mathbb{E} \left[ \left\| \mathbf{E}_\text{est}^{(b-1)} \right\|_F^2 \right] }
                               { \mathbb{E} \left[ \left\| \mathbf{Z}_\text{all}^{(b)} \right\|_F^2 \right] }.
\end{align}

This formulation enables a direct comparison with the coarse, low-rank estimate introduced in Subsection~\ref{subsec:cascate}. In particular, if the inequality
\begin{equation} 
2(1 - \rho) + \bar{\sigma}_{\rm{est}}^2 < \text{NMSE}_{\text{coarse}} 
\label{eq:comparison}
\end{equation}
is satisfied, then reusing the enhanced estimate from the previous frame offers a more accurate initialization for the iterative algorithm than relying solely on a fresh, structurally constrained raw estimate.

Assuming a LS estimator instead of LMMSE for simplicity, the value of \(\text{NMSE}_\text{coarse}\) can be computed as
\begin{equation}
\label{eq:nmse_coarse}
    \text{NMSE}_\text{coarse} = \mathbb{E}\left[\left\|\mathbf{I} - \boldsymbol{\Lambda}_p \boldsymbol{\Lambda}^{\dagger}_p\right\|^2_F\right] 
    + \frac{ \mathbb{E}\left[ \left\| \mathbf{U} \boldsymbol{\Lambda}^{\dagger}_p \right\|^2_F \right] }
           { \mathbb{E}\left[ \left\| \mathbf{Z}_\text{all}^{(b)} \right\|^2_F \right] },
\end{equation}
where \(\boldsymbol{\Lambda}_p \boldsymbol{\Lambda}^{\dagger}_p\) acts as an orthogonal projection operator onto the column space of \(\boldsymbol{\Lambda}_p\), whose rank depends on the propagation condition, given by
\begin{equation}
    \text{Rank}(\boldsymbol{\Lambda}_p) =
    \begin{cases}
    N_p/2, & \text{for LOS},\\[3pt]
    N_p, & \text{for NLOS}.
    \end{cases}
\end{equation}

Exploiting this property and the fact that \(\boldsymbol{\Lambda}_p\) is pseudo-orthogonal, (\ref{eq:nmse_coarse}) can be rewritten as
\begin{equation}
\label{eq:nmse_coarse2}
    \text{NSME}_\text{coarse} = 1 -\frac{\text{Rank}(\boldsymbol{\Lambda}_p)}{KL_e}+ \frac{\sigma_n^2}{2\sigma_z^2}.
\end{equation}
Hence, if
\begin{equation}
\label{eq:nmse_coarse3}
\color{black}
    \rho>0.5+\frac{\text{Rank}(\boldsymbol{\Lambda}_p)}{2KL_e} + \frac{\bar{\sigma}_\text{est}^2}{2}-\frac{\sigma_n^2}{4\sigma_z^2},
\end{equation}
it becomes more advantageous to use a prior estimate from a previous block—refined through the ICCE—than to rely on a coarse initial estimate.
{
In a practical NLOS scenario, for typical parameters \mbox{$N_p = 96$}, $K = 4$, $L_e = 32$, $\bar{\sigma}_\text{est}^2 \in [10^{-3}, 10^{-1}]$, and $\sigma_n^2/\sigma_z^2 \in [0,1]$, the right-hand side of (\ref{eq:nmse_coarse3}) is approximately $0.63$–$0.88$. Hence, assuming block durations on the order of tens of microseconds at $f_c=5$ GHz , this corresponds to $\rho \approx 0.99$, which is equivalent to coherence times of approximately 3–4 ms for user speeds of 6–9 m/s. Under these conditions, consecutive channel realizations remain highly correlated, confirming that the proposed reuse of the previous block’s refined estimate is indeed advantageous. This strategy not only improves accuracy but can also reduce the number of iterations required for the NMSE to converge to that of a full-pilot benchmark, improving convergence speed and efficiency.
}

Alternatively, one may prioritize reducing pilot overhead in subsequent frames by relying on coarser updates, acknowledging that this may increase $\bar{\sigma}_e^2$ and lead to some performance loss. To mitigate the accumulation of residual error, the channel estimation can be scheduled periodically, for instance, every $N_f$ frames.


We propose a two-stage channel tracking protocol for both LOS and NLOS scenarios. In the \textbf{First Stage}, the steps from Section~\ref{sec:proposed} are executed, treating channel estimation as an iterative process refined over time.
In the \textbf{Second Stage}, temporal coherence is exploited and divided into two cases, as shown in Fig.~\ref{fig:protocol}:

\begin{itemize} 
    \item \textbf{LOS:} Only the reflected channel is iteratively estimated; pilot symbols for the direct channel are still needed but remain uncoded. Specifically, $N_p'$ pilots are sent before the frame block, and the direct channel is estimated using the LMMSE estimator in (\ref{eq:lmmse}). For the reflected channel, the prior estimate is reused {($\mathbf{\hat{Z}}_\text{all}^{(b)} = \mathbf{\hat{Z}}_\text{all}^{(b-1)}$)}, followed by the iterative steps from Section~\ref{sec:proposedInt}.
    \item \textbf{NLOS:} With no direct link, the always-on direct channel protocol from Section~\ref{subsec:direct} is unnecessary. A reduced number of pilots, $N_p'' \ll N_p$, is used. As in LOS, the previous estimate is reused ($\mathbf{\hat{Z}}_\text{all}^{(b)} = \mathbf{\hat{Z}}_\text{all}^{(b-1)}$), and the encoded pilots, denoted by $N_p''$, are retained solely for SIC during decoding.
\end{itemize}

%% file: sections/Analysis.tex
\newpage
\section{Analysis}
In this section, we present an analytical investigation of the ICCE and ICT algorithms with respect to four key performance metrics: spectral efficiency, NMSE, convergence behavior, and computational complexity. Each aspect is examined to provide a comprehensive understanding of the trade-offs and benefits offered by the proposed algorithms.

\subsection{Comparison of Code Rates}
In the previous sections, we presented the ICEDD scheme which  employs encoded pilots, where pilot bits are jointly encoded with the data bits. In this approach, the number of pilots can vary from block to block, adapting to the level of temporal correlation in the channel. While this strategy enhances estimation accuracy, it introduces additional overhead and thus reduces the spectral efficiency. For a fair comparison between schemes, we define the effective code rate as
\begin{equation}
\label{eq:Reff}
    R_c^\text{eff} = \frac{\text{Number of data bits}}{\text{Total Block Length}} =  \frac{N_{b,1} + (N_f-1)N_{b,2}}{N_f T}
\end{equation}
where \(N_{b,i}\) denotes the number of data bits in Stage \(i\), \(T\) is the block length (including parity, pilots and data bits) of each frame, and \(N_f\) is the update period for channel estimation (i.e., one block is used for estimation, while the subsequent \(N_f - 1\) blocks rely on channel tracking).
 For a system without channel tracking, we set $N_f=1$ and consequently $R_c^\text{eff}=N_{b,1}/T$. 

\begin{figure}
    \centerline{\includegraphics[width=0.45\textwidth]{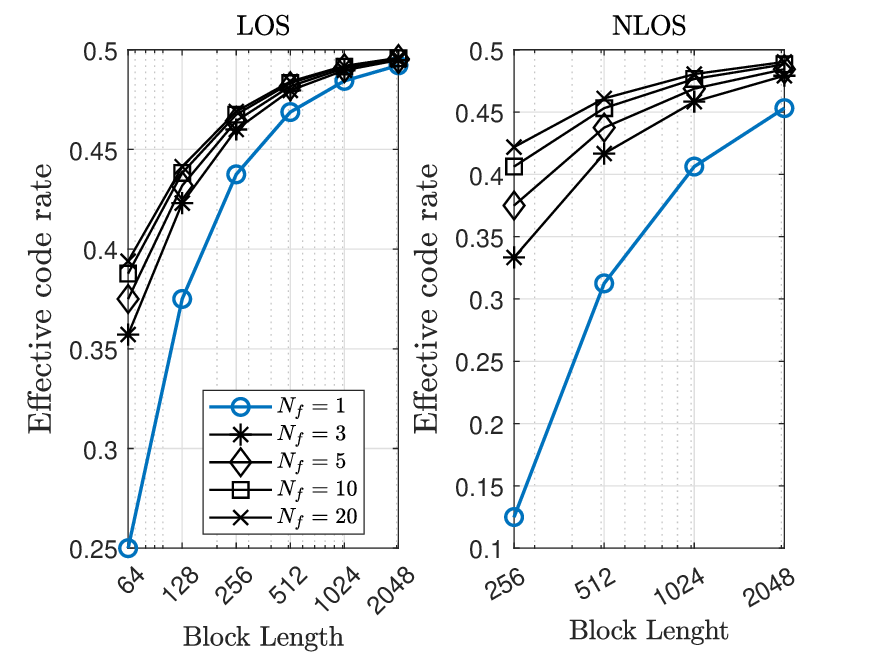}}
    \vspace{-0.75em}
    \caption{Spectral efficiency comparison.}
    \label{fig:spectral_eff}
    \vspace{-1.2em}
\end{figure}

This metric characterizes the effective payload of the system by accounting for both coding redundancy and pilot overhead within each transmission block. Based on the parameters listed in Table~\ref{tab:scenario_parameters}, the effective code rate computed using~\eqref{eq:Reff} is depicted in Fig.~\ref{fig:spectral_eff} as a function of the block length, where the number of frames $N_f$ ranges from 1 to 20. The configuration without channel tracking is indicated by the blue curve ($N_f=1$). Due to the increased pilot overhead, the NLOS scenario yields a more pronounced gain in spectral efficiency. As the number of frames \( N_f \) and block length \(N_\text{block}\) increases, both methods approach the LDPC code rate.

\begin{table}[h]
\centering
\caption{Simulation Parameters for LOS and NLOS Scenarios}
\vspace{-0.5em}
\label{tab:scenario_parameters}
\begin{tabular}{lcc}
\toprule
\textbf{Parameter}     & \textbf{LOS} & \textbf{NLOS} \\
\midrule
LDPC Code Rate             & 0.5          & 0.5          \\
Number of pilots Stage 1         & 16           & 96           \\
Number of pilots Stage 2          & 16           & 16           \\
\bottomrule
\end{tabular}
\end{table}

\subsection{MSE Analysis of Channel Estimation Error}

\textcolor{black}{
In this subsection, we derive the MSE of the ICEDD scheme for a multi-user multi-antenna system. We start by analyzing the NLOS case because, in ICEDD, the iterative refinement acts only on the cascaded user–RIS–AP channel, whereas the direct-link channel $\mathbf{H}$ is estimated independently and does not affect the computation of $\hat{\mathbf{Z}}_{\text{all,coarse}}$. For further refinements of $\hat{\mathbf{Z}}_{\text{all}}$, the presence of a direct channel introduces additional MSE components, which are analyzed later in this subsection.}


\textcolor{black}{To evaluate the impact of decoding errors on this estimation process, we characterize how symbol-level errors affect the refinement of the cascaded channel. In the proposed ICEDD scheme, all re-modulated symbols are reused as pilots, regardless of whether they were correctly decoded or not.} Assuming that the system employs QPSK modulation\footnote{The derivation can be extended to an $M$-ary constellation with $M>4$; QPSK is adopted to simplify the analytical expressions.} and that the LDPC decoder outputs binary data, the re-modulation and re-encoding process shown in Fig.~\ref{fig:blockdiagram} may introduce symbol errors due to bit-level decision errors. However, since QPSK has only four constellation points, each re-modulation error leads to one of three possible incorrect symbols. Thus, the re-modulation error can be modeled as a discrete random variable representing the phase deviation $\alpha_\Delta$ between the transmitted and decoded symbols.

\begin{table}[h!]
\centering
\caption{Phase Error Mapping in Reencoded QPSK Symbols}
\vspace{-0.5em}
\label{tab:error_analysis}
\begin{tabular}{|c|c|c|c|}
\hline
 $\alpha_\Delta $  & \( \measuredangle[{\alpha_\Delta}] \) & Error Probability ($P_b$) & Bit Errors ($N_b$) \\
\hline
\( 1 \)  & 0º     & $(1-\epsilon_b)^2$     & 0 \\
\( j \)  & 90º     & $\epsilon_b(1-\epsilon_b)$      & 1 \\
\( -1 \) & 180º    & $\epsilon_b^2$         & 2 \\
\( -j \) & 270º    & $\epsilon_b(1-\epsilon_b)$      & 1 \\
\hline
\end{tabular}
\end{table}

Assuming Gray mapping, Table~\ref{tab:error_analysis} summarizes the possible phase errors and their associated probabilities, based on the bit error probability $\epsilon_b$ for each BPSK component (in-phase and quadrature). The bit errors are assumed to be independent.

This phase distortion can be modeled by the complex-valued random variable $\alpha_\Delta$, for which the following approximations hold
\begin{align}
\label{eq:expectedValue}
\mathbb{E}\left[{\alpha_\Delta}\right] &= 1 - 2\epsilon_b, \\
\mathrm{Var}\left[{\alpha_\Delta}\right] &= 4\epsilon_b(1 - \epsilon_b) \approx 4\epsilon_b \quad \text{for } \epsilon_b \ll 1.
\end{align}

Furthermore, the average bit error rate (BER) for the QPSK system is given by the expected number of bit errors per symbol
\begin{equation}
\text{BER} = \sum P_b N_b = 2\epsilon_b(1 - \epsilon_b) + 2\epsilon_b^2 = 2\epsilon_b,
\end{equation}
which leads to the relation
\begin{equation}
\label{eq:p_value}
\epsilon_b = \frac{\text{BER}}{2}.
\end{equation}

\textcolor{black}{To incorporate decoding errors into the channel estimation model, we define a modified pilot matrix. The transmission block has a total length of $T = N_{\text{ps}} + N_p + N_{\text{info}}$, where $N_{\text{ps}}$, $N_p$, and $N_{\text{info}}$ denote the numbers of parity, pilot, and information symbols, respectively, as defined previously. Each column of the ideal matrix $ \boldsymbol{\Lambda} \in \mathbb{C}^{KL_e \times T} $, as introduced in~(\ref{eq:cascade_2}) is distorted by a diagonal matrix $ \boldsymbol{\Omega}^{(i)} \in \mathbb{C}^{K \times K} $. The entries of $ \boldsymbol{\Omega}^{(i)} $ are independent random variables whose values and associated probabilities are defined in Table~\ref{tab:error_analysis}. The resulting reference matrix, $ \boldsymbol{\Lambda}_\text{ref} \in \mathbb{C}^{KL_e \times T} $, is then given by}
\begin{equation}
    \boldsymbol{\Lambda}_\text{ref} = \left[(\boldsymbol{\Omega}^{(1)}\mathbf{x}^{(1)}) \otimes \boldsymbol{\varphi}^{(1)} \;\cdots\; (\boldsymbol{\Omega}^{(T)}\boldsymbol{x}^{(T)}) \otimes \boldsymbol{\varphi}^{(T)}\right],
\end{equation}
where
\begin{align}
    \boldsymbol{\Omega}^{(i)} \mathbf{x}^{(i)} &=
    \begin{bmatrix}
    \Delta\alpha_{1}^{(i)} & \cdots & 0 \\
    \vdots & \ddots & \vdots \\
    0 & \cdots & \Delta\alpha_{K}^{(i)}
    \end{bmatrix}
    \begin{bmatrix}
    x_1^{(i)} \\
    \vdots \\
    x_K^{(i)}
    \end{bmatrix}
    =
    \begin{bmatrix}
    \Delta\alpha_{1}^{(i)} x_1^{(i)} \\
    \vdots \\
    \Delta\alpha_{K}^{(i)} x_K^{(i)}
    \end{bmatrix}.
\end{align}

Each column of \( \boldsymbol{\Lambda}_\text{ref} \) can also be expressed using the Kronecker product:
\begin{align}
     (\boldsymbol{\Omega}^{(i)} \mathbf{x}^{(i)}) \otimes \boldsymbol{\varphi}^{(i)} 
     &= (\boldsymbol{\Omega}^{(i)} \otimes \mathbf{I}_{L_e})(\mathbf{x}^{(i)} \otimes \boldsymbol{\varphi}^{(i)}) \nonumber \\
     &= (\boldsymbol{\Omega}^{(i)} \otimes \mathbf{I}_{L_e})\boldsymbol{\lambda}^{(i)}.
\end{align}

Using the \textcolor{black}{reference matrix}, the LS estimator from (\ref{eq:cascade_2}) becomes:
\begin{equation}
    \label{eq:ls_estimator}
    \mathbf{\hat{Z}}_\text{all} = \mathbf{Y}_\text{cascade} \boldsymbol{\Lambda}_\text{ref}^\dagger 
    = \mathbf{Z}_\text{all} \boldsymbol{\Lambda} \boldsymbol{\Lambda}_\text{ref}^\dagger 
    + \mathbf{U} \boldsymbol{\Lambda}_\text{ref}^\dagger.
\end{equation}

Defining \( \boldsymbol{\Omega}^{(i)}_\text{kron} = \boldsymbol{\Omega}^{(i)} \otimes \mathbf{I}_N \), which remains diagonal and unitary, we can expand the product \( \boldsymbol{\Lambda} \boldsymbol{\Lambda}_\text{ref}^\dagger \), where both \( \boldsymbol{\Omega}^{(i)}_\text{kron} \) and \( \boldsymbol{\Lambda} \boldsymbol{\Lambda}_\text{ref}^\dagger \) are complex matrices in \( \mathbb{C}^{KL_e \times KL_e} \), as
\begin{flalign}
\label{eq:lamblamb}
\boldsymbol{\Lambda} \boldsymbol{\Lambda}_\text{ref}^\dagger 
&= 
\left[
\sum_{i=1}^{T} (\boldsymbol{\Omega}^{(i)}_\text{kron})^H \boldsymbol{\lambda}^{(i)} (\boldsymbol{\lambda}^{(i)})^H 
\right]
\left[
\sum_{i=1}^{T} \boldsymbol{\lambda}^{(i)} (\boldsymbol{\lambda}^{(i)})^H 
\right]^{-1}.
\end{flalign}

Assuming sufficiently large \( T \), we approximate the second sum by its expectation
\begin{equation}
\label{eq:approx}
 \sum_{i=1}^{T} \boldsymbol{\lambda}^{(i)} (\boldsymbol{\lambda}^{(i)})^H 
\approx \mathbb{E}[\boldsymbol{\lambda}^{(i)} (\boldsymbol{\lambda}^{(i)})^H] T = \sigma_x^2 T \mathbf{I}_{KL_e}.
\end{equation}

Substituting into (\ref{eq:lamblamb}), we get
\begin{equation}
\label{eq:lamblambH}
\boldsymbol{\Lambda} \boldsymbol{\Lambda}_\text{ref}^\dagger 
\approx \frac{1}{\sigma_x^2 T} \sum_{i=1}^{T} (\boldsymbol{\Omega}^{(i)}_\text{kron})^H \boldsymbol{\lambda}^{(i)} (\boldsymbol{\lambda}^{(i)})^H.
\end{equation}

Using this approximation in the LS estimate (\ref{eq:ls_estimator}), we write the MSE as
\begin{align}
\label{eq:mse}
\text{MSE} &= \mathbb{E}\left[\left\| \widehat{\mathbf{Z}}_{\text{all}} - \mathbf{Z}_{\text{all}} \right\|_F^2 \right] \nonumber \\ 
&= \underbrace{\mathbb{E}\left[\left\| \mathbf{Z}_{\text{all}} \left( \boldsymbol{\Lambda} \boldsymbol{\Lambda}_{\text{ref}}^\dagger - \mathbf{I} \right) \right\|_F^2 \right]}_{\text{MSE}_1}
+ \underbrace{\mathbb{E}\left[\left\| \mathbf{U} \boldsymbol{\Lambda}_{\text{ref}}^\dagger \right\|_F^2 \right]}_{\text{MSE}_2}.
\end{align}

Assuming uncorrelated entries in \( \mathbf{Z}_\text{all} \sim \mathcal{CN}(0, \sigma_z^2) \), the first term becomes
\begin{align}
\label{eq:mse_1}
\text{MSE}_1 
&= M \sigma_z^2 \cdot \text{Tr} \left\{ 
\mathbb{E} \left[ 
\left( \boldsymbol{\Lambda} \boldsymbol{\Lambda}_{\text{ref}}^\dagger - \mathbf{I} \right)^H 
\left( \boldsymbol{\Lambda} \boldsymbol{\Lambda}_{\text{ref}}^\dagger - \mathbf{I} \right) 
\right] 
\right\} \nonumber \\
&= M \sigma_z^2 \cdot \text{Tr} \left\{ 
\mathbb{E} \left[ \left\| \boldsymbol{\Lambda} \boldsymbol{\Lambda}_{\text{ref}}^\dagger \right\|^2 \right] 
- 2 \mathbb{E} \left[ \boldsymbol{\Lambda} \boldsymbol{\Lambda}_{\text{ref}}^\dagger \right] 
+ \mathbf{I} 
\right\} \nonumber \\
&={\mathbb{E}\left[\|\mathbf{Z}_{\mathrm{all}}\|_F^2\right]}\left[\frac{N(K-(1-2p)^2)}{T} +4p^2\right],
\end{align}
where the derivation of \(\text{MSE}_1\) is provided in Appendix~\ref{app:lambda}.

For the second term, the noise term, we use results from random matrix theory and the Wishart approximation~\cite{tulino2004random} for \( \boldsymbol{\Lambda}_{\text{ref}} \left( \boldsymbol{\Lambda}_{\text{ref}} \right)^H \) to write
\begin{align}
\label{eq:mse_2}
\text{MSE}_2 
&= \mathbb{E}\left[\left\| \mathbf{U} \boldsymbol{\Lambda}_{\text{ref}}^\dagger \right\|_F^2 \right] \nonumber \\
&= M \sigma_n^2 \cdot \text{Tr} \left[ \left( \boldsymbol{\Lambda}_{\text{ref}} \left( \boldsymbol{\Lambda}_{\text{ref}} \right)^H \right)^{-1} \right] \nonumber \\
&= \frac{KML_e}{(T - KL_e) \sigma_x^2} \cdot \sigma_n^2
\end{align}
for $T > KL_e$.

Substituting \eqref{eq:mse_1} and \eqref{eq:mse_2} into \eqref{eq:mse} and normalizing, the NMSE (NMSE $= \text{MSE}/{\mathbb{E}\left[\|\mathbf{Z}_{\mathrm{all}}\|_F^2\right]} $) is given by
\begin{equation}
    \label{eq:nmse_nlos}
    \mathrm{NMSE}_\mathrm{NLOS}  
    = \frac{L_e(K-(1-2p)^2)}{T} + \frac{\sigma_n^2}{(T - KL_e) \sigma_x^2 \sigma_z^2}+4p^2.
\end{equation}

From this expression, it follows that in the absence of systematic errors ($p \rightarrow 0$), and for sufficiently large pilot length for~\eqref{eq:approx} to hold (i.e., $T \gg L_eK$), the NMSE attains its minimum value:
\begin{equation}
    \mathrm{NMSE}_{\text{NLOS},\min} \approx  \frac{\sigma_n^2}{(T - KL_e) \sigma_x^2 \sigma_z^2}.
\end{equation}

{
The MSE analysis for the iterative estimation process can be further extended to scenarios involving a LOS link. Let $e' \triangleq \mathrm{NMSE}_{\mathrm{LOS}}$. The derivation shown in Appendix~\ref{app:B} yields
\begin{equation}
e'
=\kappa\!\left[\frac{L_e(K\!-\!(1\!-\!2p)^2)}{T}
+\frac{\sigma_n^2\varrho_e}{(T\!-\!KL_e)\sigma_x^2\sigma_z^2}
+4p^2\right]
\end{equation}
where
\begin{equation}
\varrho_e = \left(\frac{K\sigma_h^2}{\sigma_n^2 + 2K\sigma_x^2\sigma_h^2} + 1\right) \;\;\text{and}\;\; \kappa = \frac{L_e\sigma_z^2}{\sigma_h^2+L_e\sigma_z^2}.
\end{equation}

It is worth noting that when $\sigma_h^2 \rightarrow 0$, the direct link vanishes and the expression naturally reduces to the NLOS case, confirming the consistency of the model. 
Conversely, for large $\sigma_h^2$, corresponding to a strong LOS component, the scaling factor $\kappa = \frac{L_e\sigma_z^2}{\sigma_h^2 + L_e\sigma_z^2}$ decreases approximately as $1/\sigma_h^2$, while $\varrho_e$ saturates to a finite value. 
As a result, $e'$ also decreases proportionally to $1/\sigma_h^2$. 
This implies that, in strongly LOS-dominated scenarios, the estimation error associated with the cascaded channel can be largely reduced, and the overall NMSE decreases accordingly.
  
}
\subsection{Convergence and SNR Threshold}
In this subsection, we derive the evolution of the NMSE for channel estimation as a function of the number of iterations and compute the SNR necessary for the iterative method to converge. The channel estimation error can be expressed as
\begin{equation}
    \mathbf{Z}_1 = \mathbf{\hat{Z}}_1 + \mathbf{E}_{\text{est}}^{(\beta)},
    \label{eq:estimation error}
\end{equation}
where $\mathbf{E}_{\text{est}}^{(\beta)}$ denotes the channel estimation error at iteration $\beta$ of the ICCE, where the channel estimation is performed/enhanced. {Approximating this error by the matrix $\mathbf{E}_{\text{est}}^{(\beta)} \sim \mathcal{CN}(\mathbf{0}_M, \sigma_{e,\beta}^2M\mathbf{I}_{L_e})$ with complex Gaussian random variables.} Substituting \eqref{eq:estimation error} into (\ref{eq:eq04}) and assuming a NLOS channel, we obtain
\begin{align}
    \mathbf{y}^{(i)}    &=\sum_{j=1}^K(\mathbf{\hat{Z}}_j+\mathbf{E}_{\text{est},j}^{(\beta)})\boldsymbol{\varphi}_e^{(i)}x_j^{(i)} + 
     \mathbf{n}^{(i)}  \\
    &= \mathbf{\hat{Z}}_k \boldsymbol{\varphi}^{(i)} x^{(i)}_k +\sum_{j \neq k}^{K}\mathbf{\hat{Z}}_j\boldsymbol{\varphi}_e^{(i)}x_j^{(i)} + \sum_{j = 1}^{K}\mathbf{E}_{\text{est},j}^{(\beta)}\boldsymbol{\varphi}_e^{(i)}x_j^{(i)} + \mathbf{n}^{(i)} \nonumber
\end{align}

Assuming that each reflective element applies an independent random phase shift, representing a suboptimal receiver scenario (i.e., worse than coherent combining as discussed in \cite{9998527}), the multiplication by \( \boldsymbol{\varphi}^{(i)}_e \) introduces a random phase rotation. Then, the terms \( \mathbf{\hat{Z}}_j \boldsymbol{\varphi}_e^{(i)} \) and \( \mathbf{E}_{\text{est}}^{(\beta)} \boldsymbol{\varphi}_e^{(i)} \) become linear combinations of \( M \) independent Gaussian variables. Therefore, they follow complex Gaussian distributions given by
\begin{equation}
     \boldsymbol{\hat{\upsilon}}_j = \mathbf{\hat{Z}}_j \boldsymbol{\varphi}_e^{(i)} \sim \mathcal{CN}(\mathbf{0},\sigma_z^2 M \mathbf{I}_{L_e}),
\end{equation}
\begin{equation}
     \boldsymbol{\xi }_j^{(\beta)} = \mathbf{E}_{\text{est}}^{(\beta)} \boldsymbol{\varphi}_e^{(i)} \sim \mathcal{CN}(\mathbf{0},\sigma_{e,\beta}^2 M \mathbf{I}_{L_e}).
\end{equation}

Defining the estimated channel matrix as \( \Upsilon = [\boldsymbol{\hat{\upsilon}}_1, \dots, \boldsymbol{\hat{\upsilon}}_k]^T \), and applying a linear receive filter as in~\eqref{detection_estimate_1}, the detected symbol \( \hat{x}_k \) can be expressed as the sum of three components: signal plus interference, estimation noise, and AWGN noise, as follows:
\begin{equation}
    \color{black}
    \hat{x}_k = \underbrace{(\mathbf{w}_k^{(i)})^H \Upsilon \mathbf{x}^{(i)}}_{\text{signal + interference}} 
    + \underbrace{(\mathbf{w}_k^{(i)})^H \boldsymbol{\xi}_j^{(\beta)} \mathbf{x}^{(i)}}_{\text{estimation noise}} 
    + \underbrace{(\mathbf{w}_k^{(i)})^H \mathbf{n}^{(i)}}_{\text{AWGN noise}}.
\end{equation}
With this representation we can approximate the results for the linear detector in \cite{8917985}, where the SINR follows a chi-square distribution, where the mean SINR per channel is given by
\begin{equation}
    \bar{\gamma}  =
   \frac{M\sigma_z^2\sigma_x^2}{ M\sigma_{e,\beta}^2 \sigma_x^2 + \sigma_n^2}= \frac{\sigma_z^2}{\sigma_{e,\beta}^2 + \frac{\sigma_n^2}{M\sigma_x^2}} 
\end{equation}
Assuming that $\sigma_{e,\beta}^2 \gg \frac{\sigma_n^2}{M\sigma_x^2}$, the mean SINR can be approximated by $\bar{\gamma} \approx{\sigma_z^2}/{\sigma_{e,\beta}^2}$. Using this SINR as SNR (negligible interference), the analytical expression for the bit error rate (BER) with QPSK modulation can be approximated by
\begin{equation}
    \text{BER}_\text{QPSK} \approx Q\left( \sqrt{\frac{\sigma_z^2}{\sigma_{e,\beta}^2}G_c} \right),
\end{equation}
where $G_c$ denotes the coding gain, with $G_c = 1$ for an uncoded system.

Now, comparing the estimation error at iteration $\beta$ from (\ref{eq:estimation error}) with that at iteration $(\beta+1)$ from (\ref{eq:mse}), and substituting $p$ from (\ref{eq:p_value}), we obtain
\begin{align}
    \label{eq:mse_rho}
    \text{NMSE}^{(\beta)} &\approx {\sigma_{e,\beta}^2} \\
    \text{NMSE}^{(\beta+1)} &\approx \frac{L_e(K-(1-2p)^2)}{T}  \nonumber \\
    &+ \frac{\sigma_n^2}{(T - KL_e) \sigma_x^2 \sigma_z^2}+Q\left( \sqrt{\frac{\sigma_z^2}{\sigma_{e,\beta}^2}G_c} \right)^2.
    \label{eq:mse_rho2}
\end{align}

Considering a sufficiently large pilot length such that~\eqref{eq:approx} holds (i.e., \( T \gg L_eK \)) and noting that \( (1 - 2p)^2 \le 1 \), the term \( \frac{L_e(K - (1 - 2p)^2)}{T} \) tends to zero. Therefore, we have
\begin{equation}
    \text{NMSE}^{(\beta+1)} = \text{NMSE}_\text{min} + Q\left( \sqrt{\frac{\sigma_z^2}{\sigma_{e,\beta}^2}G_c} \right)^2.  
\end{equation}

Intuitively, as the estimation error variance ${\text{NMSE}^{(\beta)}}$ $({\sigma_{e,\beta}^2})$ decreases, the Q-function argument increases, leading to a smaller value of $Q(\cdot)$ and hence a lower ${\text{NMSE}^{(\beta+1)}}$. When $\sigma_e^2 \to 0$, the Q-function tends to zero (i.e., all symbols are correctly decoded), and the minimum is achievable. This represents the performance limit of the ICEDD scheme. 

For the convergence of the estimation enhancement, the NMSE must decrease with each iteration, by comparing equations (\ref{eq:mse_rho}) and (\ref{eq:mse_rho2}), we arrive at the following condition:

\begin{align}
   &\text{NMSE}^{(\beta)} > \text{NMSE}^{(\beta+1)},  \\
   & \sigma_{e,\beta}^2 > \text{NMSE}_\text{min} + Q\left( \sqrt{\frac{\sigma_z^2}{\sigma_{e,\beta}^2}G_c} \right)^2, \\
   & \sigma_{e,\beta}^2 -Q\left( \sqrt{\frac{\sigma_z^2}{\sigma_{e,\beta}^2}G_c} \right)^2 > \frac{\sigma_n^2}{(T - KL_e) \sigma_x^2 \sigma_z^2}\\
   &\frac{\sigma_x^2}{\sigma_n^2}>\frac{1}{(T - KL_e) \left(\sigma_{e,\beta}^2 -Q\left( \sqrt{\frac{\sigma_z^2}{\sigma_{e,\beta}^2}G_c} \right)^2\right) \sigma_z^2}
\end{align}

From the inequality above, we observe that the required user transmit power $\sigma_x^2$ must exceed a certain threshold to ensure the convergence of the iterative processing. This threshold is influenced by the initial channel estimation error $\sigma_e^2$, the variance of the effective channel $\sigma_{z}^2$, and the coding gain $G_c$.

The estimation error $\sigma_e^2$ is inversely related to the number of pilot symbols used in the coarse (initial) channel estimation. A smaller number of pilots leads to a higher estimation error, which in turn increases the minimum required transmit power for convergence. Conversely, using more pilots improves the initial estimate, reducing the power required for the enhancement process to be effective.
Furthermore, the inequality reveals two additional key insights:
\begin{itemize}
    \item A higher coding gain $G_c$ improves convergence conditions by reducing the required transmit power.
    \item An increased channel variance $\sigma_{z}^2$ also reduces the power needed for convergence.
\end{itemize}
These observations highlight the impact of pilot allocation, transmission power, coding strategies, and channel on the performance of the ICCE algorithm.

\subsection{Computational Complexity}
{The computational complexity of the ICCE and ICT algorithms is evaluated in terms of the number of complex floating-point operations (flops) required for the IDD and channel estimation. A simplified system model, shown in Fig.~\ref{fig:complexity_bd}, is adopted for the analysis. The exact cost may vary slightly depending on the specific implementation, and it is assumed that $N > M > K$.

    \begin{figure}[h!]
        \centerline{\includegraphics[width=0.4\textwidth]{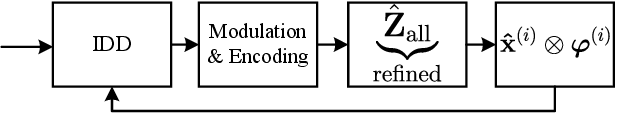}}
        \caption{Simplified system model.}
        \label{fig:complexity_bd}
    \end{figure}

\begin{enumerate}
    \item \textbf{IDD block:}
    The dominant term stems from the matrix inversion in (\ref{eq:w}). Using Cholesky factorization \cite{1494998}, the complexity is $\mathcal{O}\left(\tfrac{2}{3}M^3\right)$. This operation is performed for $K$ users and $T$ symbols in each packet, at every IDD iteration.
    The LDPC decoding complexity per iteration is approximately $\mathcal{O}(20T\varsigma)$, where $\varsigma$ is the number of decoding iterations. In this work, $\varsigma = 10$, which is relatively small compared to the matrix operations. Therefore, the overall dominant term is
    \begin{equation*}
    \mathcal{O}\left(\tfrac{2}{3}M^3 K T \mu \right),
    \end{equation*}
    where $\mu$ denotes the number of detector–decoder iterations in the IDD process.
        
    \item \textbf{Modulation \& Encoding:} 
    Encoding uses a systematic generator matrix $\mathbf{G}_\mathrm{LDPC}\in\{0,1\}^{TRM_b\times TM_b}$ with code rate $R$. The coding is performed in GF(2) as $\mathbf{c}=\mathbf{b}\mathbf{G}_\mathrm{LDPC}$ for each user. The cost of the direct matrix product (bitwise AND and XOR operations) is
    \begin{equation*}
        \mathcal{O}_b\big(2\,T^2 R(1-R)M_b^2\big)\ \text{per user}.
    \end{equation*}      
    Assuming $R=1/2$ and QPSK ($M_b=2$), and considering $K$ users, this simplifies to
    \begin{equation*}
    \mathcal{O}_b(2T^2K)\ \text{total},
    \end{equation*}
    which is negligible compared to the IDD complexity.

    \item ${\hat{\mathbf{x}}^{(i)} \otimes {\boldsymbol{\varphi}}^{(i)}}$ \textbf{:} 
    Considering all transmitted symbols, the complexity is
    \begin{equation*}
    \mathcal{O}(KNT),
    \end{equation*}
    which is negligible relative to the other processing blocks.
        
    \item \textbf{Cascaded channel refinement:}  
    The main computational cost arises from (\ref{eq:Z_ref}) when computing $\hat{\mathbf{H}}_\text{LMMSE}$. Assuming $T \approx L_eK$, the dominant term is
    \begin{equation*}
        \mathcal{O}\left(\tfrac{2}{3} L_e^3 K^3 \right). 
    \end{equation*}
\end{enumerate}
    
Assuming that $\beta$ denote the ICCE iteration, the total complexity of this system can be defined as:
\begin{equation*}
    \mathcal{O}\!\left(\frac{\left[M^3 K T \mu + L_e^3 K^3\right]2\beta}{3}\right)
\end{equation*}
}

%% file: sections/results.tex
 \section{Numerical Results}

In this section, we assess the proposed ICEDD scheme along with the ICCE and ICT algorithms against competing approaches using numerical simulations. A regular LDPC code, generated using the McKay method with a block length of 2048 and a code rate of 1/2, is employed alongside the QPSK modulation. The channel is assumed to experience block fading and the estimation of the channel state information (CSI) is done at the AP receiver. Two uplink scenarios were evaluated: LOS (weakened direct link with strong reflected link) and NLOS (no direct link and strong reflected link). Path loss models follow the 3GPP standard \cite{access2010further}, and the system parameters are presented in Table \ref{tab:parameters}. To account for small-scale fading effects, a Rayleigh fading channel model was adopted for all examples. Each figure displays two results for the same scenario using subplots. A shared legend is included and applies to both plots.\vspace{-0.15em}

\begin{table}[ht]
    \caption{Simulation Parameters}\vspace{-0.75em}
    \label{tab:parameters}
\centering
\begin{tabular}{c|c}
\hline
{\textbf{Parameters}}               & { \textbf{Values}} \\ \hline
Frequency                           & 5 GHz              \\ \hline
Bandwidth                           & 1 MHz              \\ \hline
Noise power spectral density        & -170 dBm/Hz              \\ \hline
Path loss AP-RIS; RIS-Users (dB)    & $37.3+22log_{10}(d)$  \\ \hline
Path loss AP-Users (dB)             & $32.4+30log_{10}(d)$  \\ \hline
Numer of AP antennas                & 8       \\ \hline
Number of Users                     & 4       \\ \hline
Number of RISs                      & 2      \\ \hline
Number of Cells (per RIS)           & 16     \\ \hline
Location of AP                      & (0 m, 0 m, 0 m)       \\ \hline
Location of RIS$_1$                 & (500 m, 10 m, 0 m)     \\ \hline
Location of RIS$_2$                 & (500 m, -10 m, 0 m)    \\ \hline
Geometric center of users positions & (500 m, 0 m, 0 m)      \\ \hline
Users Spatial Radius                & 5 m                   \\ \hline
\end{tabular}
\end{table}

\begin{figure}
    \vspace{-1em}
\includegraphics[width=0.525\textwidth]{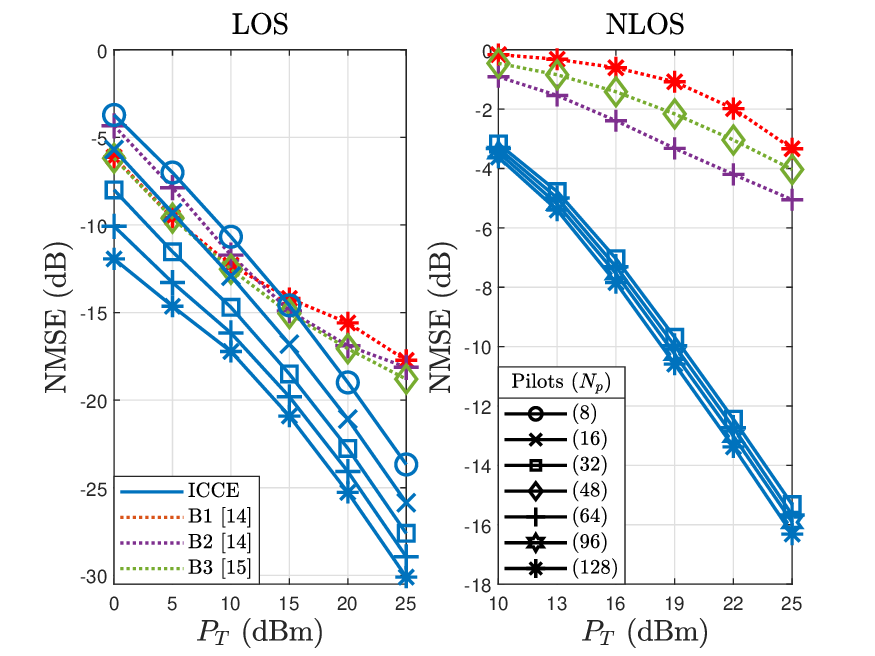}    \vspace{-2.0em}
    \caption{Comparison between ICCE and other channel estimation techniques proposed in the literature.}
    \label{fig:trans_00}
\vspace{-0.5em}
\end{figure}

In the curves presented in Fig. \ref{fig:trans_00}, we compare the Performance Limit (PL) — defined as the lowest achievable NMSE of channel estimation when all symbols are perfectly decoded at a given transmit power per user $P_T$ — of the proposed ICCED scheme with three baseline channel estimation methods: (i) Baseline 1 (B1)\cite{9130088}, which uses a conventional on–off channel estimation protocol; (ii) Baseline 2 (B2)\cite{9130088}, a three-phase pilot-based channel estimation framework that exploits the correlation among the user–IRS–BS channels; and (iii) Baseline 3 (B3)\cite{9839429}, which leverages the common-link structure to reduce pilot overhead. Fig. \ref{fig:trans_00} illustrates the NMSE performance versus the transmit power of the users $P_T$, demonstrating the effectiveness of the proposed ICEDD scheme and the superiority of the proposed ICCE algorithm to the baselines. As each technique operates with different pilot lengths, the results are shown for multiple pilot configurations.

In the LOS scenario, ICCE achieves the same estimation accuracy using only $N_p = 16$ pilots at low SNR, compared to 128, 64, and 48 pilots required by B1, B2, and B3, respectively. In the NLOS scenario, ICCE requires a larger number of pilot symbols due to the absence of a direct link. Nevertheless, it still provides a significant performance gain compared to the conventional baselines.

\begin{figure}
    \vspace{-1em}
\includegraphics[width=0.525\textwidth]{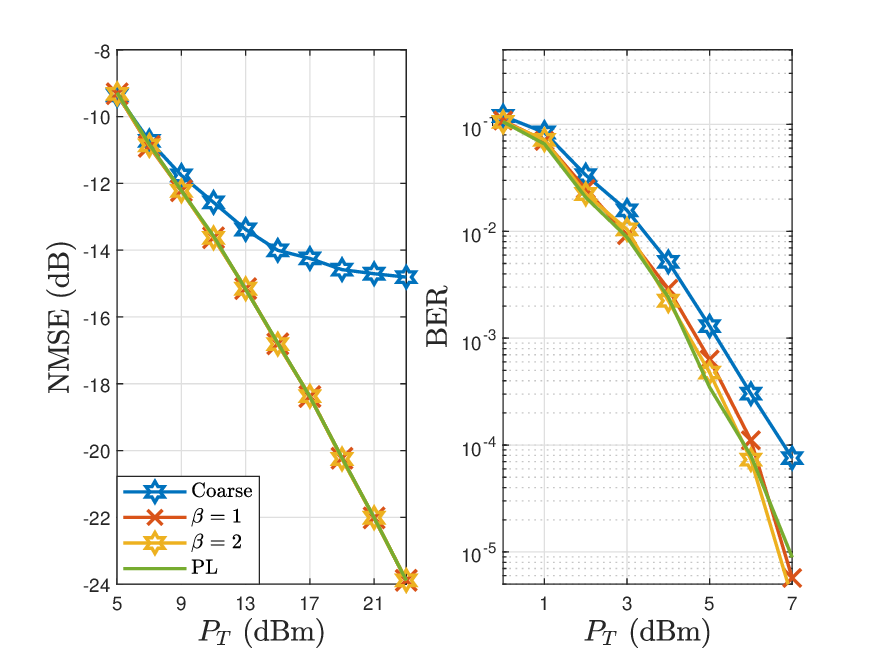}
  \vspace{-2em}
    \caption{ICCE LOS RIS-Assisted system with $N_p=16$.}
    \label{fig:trans_01}
    \vspace{-2em}
\end{figure}

\begin{figure}
    \vspace{-1em}
\includegraphics[width=0.525\textwidth]{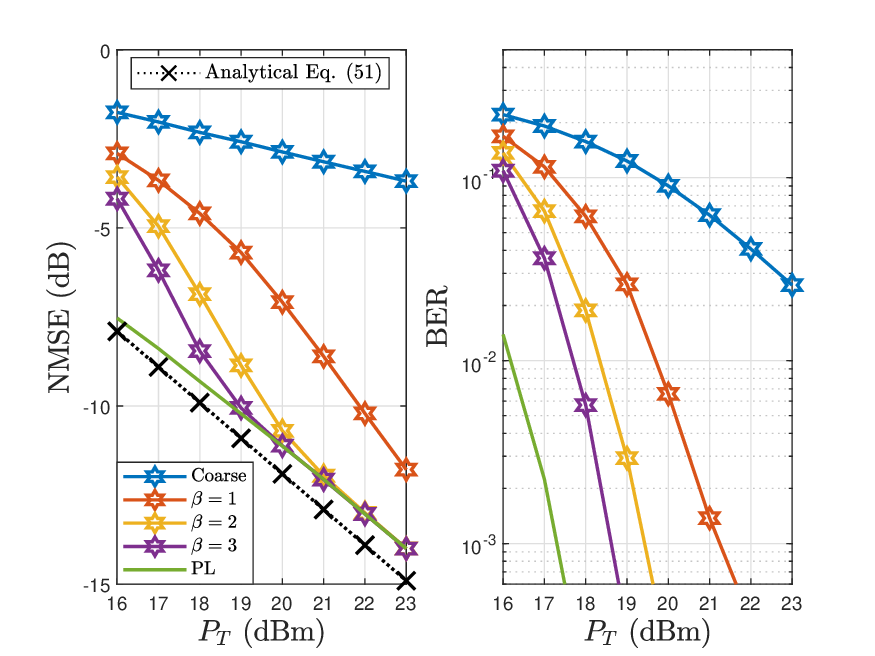}
  \vspace{-1em}
    \caption{ICCE NLOS RIS-Assisted system with $N_p=96$.}
    \label{fig:trans_02}
    \vspace{-1em}
\end{figure}

\begin{figure}
    \vspace{-1em}
\includegraphics[width=0.525\textwidth]{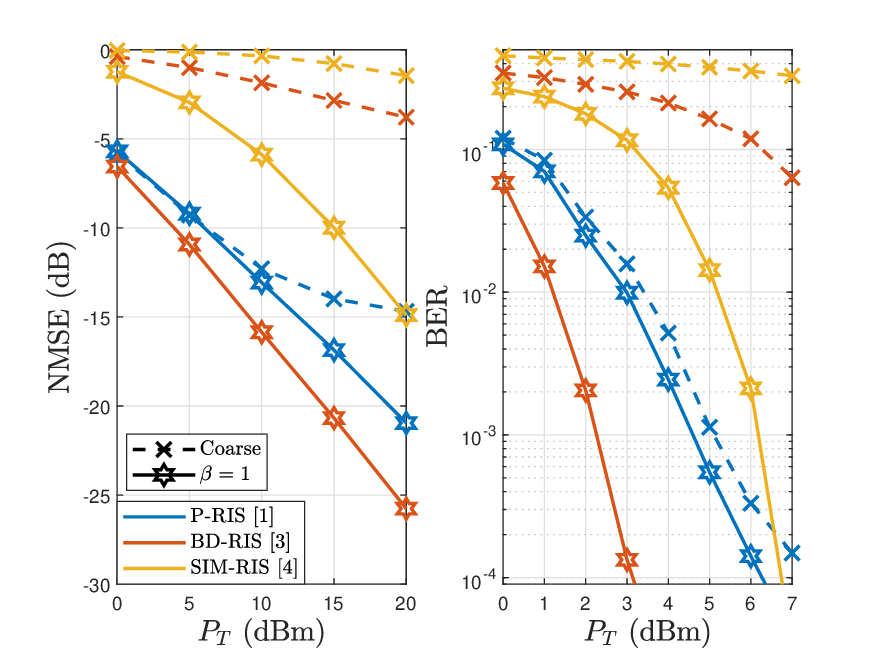}
  \vspace{-1em}
    \caption{ICCE LOS RIS-assisted system with varying extensions.}
    \label{fig:trans_03}
    \vspace{-1em}
\end{figure}

\begin{figure}
    \vspace{-1em}
\includegraphics[width=0.525\textwidth]{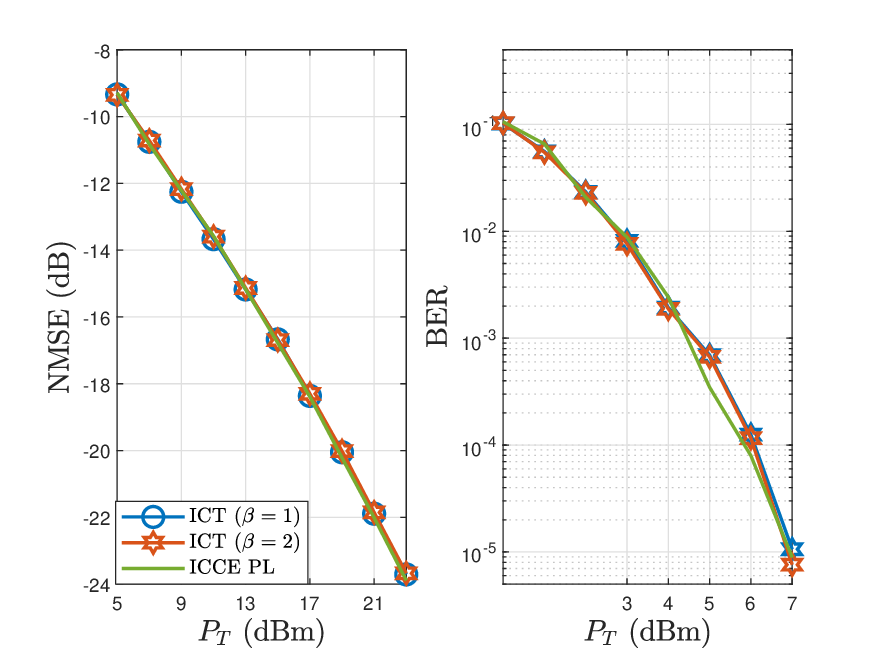}
  \vspace{-1em}
    \caption{ICT LOS RIS-Assisted system with $N_p=N_p'=16$.}
    \label{fig:trans_04}
    \vspace{-1em}
\end{figure}

\begin{figure}
    \vspace{-1em}
\includegraphics[width=0.525\textwidth]{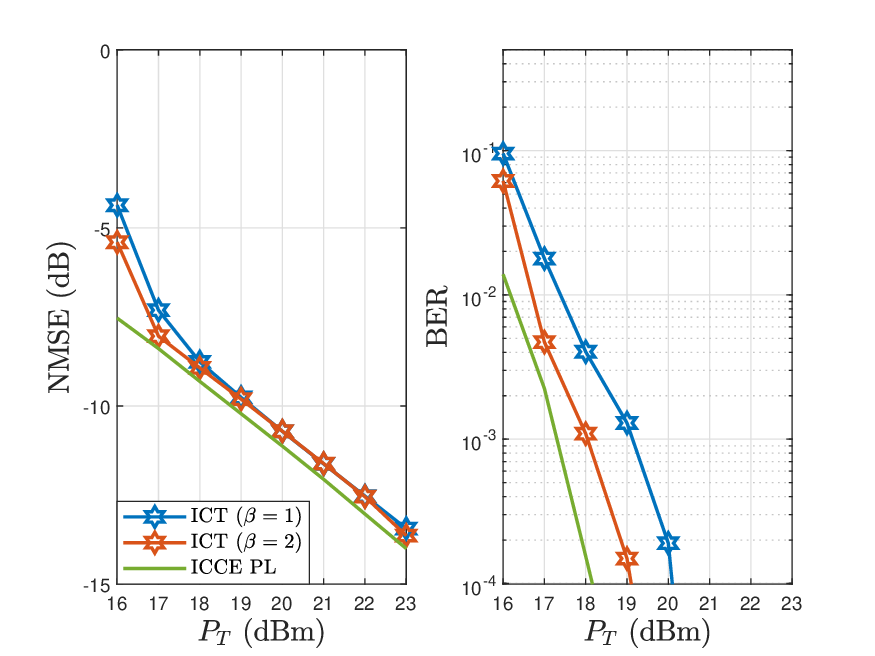}
  \vspace{-1em}
    \caption{ICT NLOS RIS-Assisted system with $N_p = 96$  and  $N_p'' = 16$.}
    \label{fig:trans_05}
    \vspace{-1em}
\end{figure}

The results of ICCE in terms of NMSE and BER as functions of $P_T$ for two specific scenarios, LOS and NLOS, are depicted in Figs.~\ref{fig:trans_01} and~\ref{fig:trans_02}, respectively. The parameter $\beta$ corresponds to the iteration index in the ICCE channel estimation process, as defined earlier. {The performance improves as the number of $P_T$ increases. 
In Figs.~\ref{fig:trans_01}, the apparent difference in performance curves for the coarse estimation is mainly due to the chosen $x$-axis scales: in the NMSE plot, $P_T$ ranges from 5 to 22~dBm, while in the BER plot it ranges from 0 to 7~dBm}. In the LOS case, the PL is achieved with only a few iterations ($\beta = 1$) due to the presence of a direct link, which significantly aids in estimating the received symbols. Conversely, in the NLOS scenario, where the direct link is absent, multiple iterations (depending on the SNR) are required to reach the peformance limit. In both scenarios, improvements in NMSE and BER are observed, particularly at medium-to-high SNR levels.

Figure~\ref{fig:trans_03} presents the results for a LOS scenario using ICCE with a block length of 1024 and various RIS configurations. The results demonstrate that ICCE can be adapted to different RIS setups and effectively exploits their characteristics. Note that the SIM-RIS yields the lowest performance, whereas the beyond-diagonal configurations achieve the best results.

Figs.~\ref{fig:trans_04} and~\ref{fig:trans_05} evaluate the ICT algorithm’s performance in LOS and NLOS scenarios, respectively, exploiting temporal coherence with the parameters in Table~\ref{tab:ICT}. {In both cases, time-correlated channel samples were generated according to the Jakes model.} In the LOS case, ICT achieves estimation performance comparable to the ICCE PL while employing channel tracking and operating at a higher effective code rate. For NLOS, shorter coded pilot sequences are used in the second stage to improve spectral efficiency. This results in slight performance degradation; however, as SNR increases, the ICT NMSE approaches the performance limit.

Comparing with the ICCE-based system in NLOS (Fig.~\ref{fig:trans_02}), ICT converges more rapidly despite the slight degradation. This gain stems from refined channel estimates leveraging temporal correlation across blocks, allowing reduced pilot overhead while maintaining higher accuracy than the initial coarse estimates.
Finally, under NLOS conditions, ICT requires more pilot symbols to compensate for reduced channel observability due to the lack of a direct path. \textcolor{black}{Although its estimation performance remains slightly below that of the ICCE PL, the gap narrows considerably at higher SNRs, highlighting the benefits of exploiting temporal coherence.}

\begin{figure}
    \vspace{-1em}
\includegraphics[width=0.525\textwidth]{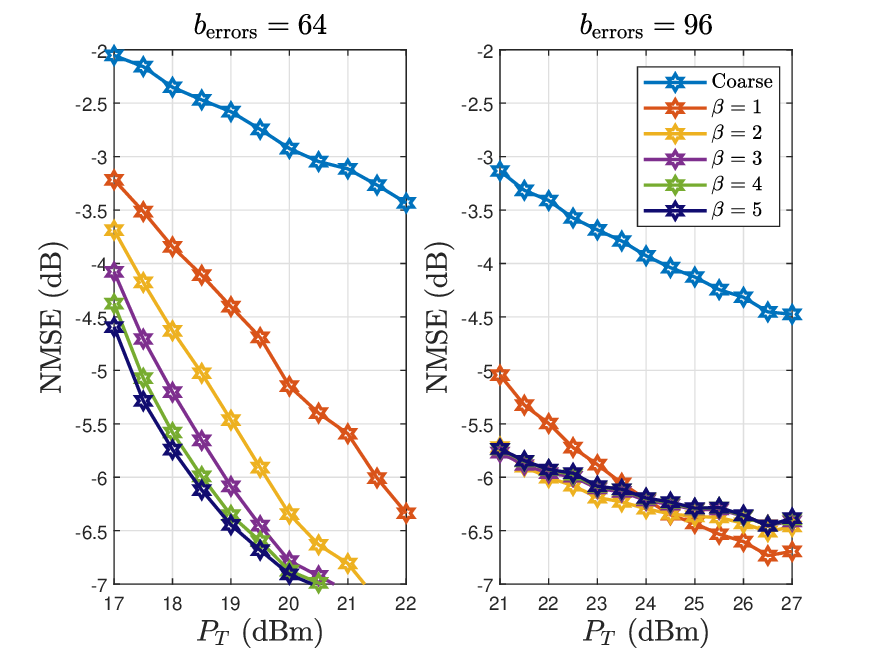}
  \vspace{-1em}
    \caption{NMSE performance under burst error injection with different burst lengths.}
    \label{fig:trans_burst}
    \vspace{-1em}
\end{figure}

{To evaluate the impact of error propagation, burst errors of different lengths were injected into the coded blocks. As illustrated in Fig.~\ref{fig:trans_burst}, the proposed framework maintains convergence for moderate burst lengths (\(b_\mathrm{errors} \le 64\)), exhibiting no substantial accumulation of estimation errors across iterations. Noticeable degradation arises only under severe burst conditions (\(b_\mathrm{errors}=96\)) combined with high transmit power levels (\(P_T > 25\)~dBm), indicating localized error propagation. Overall, these observations demonstrate that the proposed method is robust to burst-induced disturbances and preserves stable operation under typical operating scenarios.}

\vspace{-0.4cm}

\begin{table}[h]
\label{tab:ICT}
\centering
\caption{Additional parameters for ICT } \vspace{-0.75em}
\begin{tabular}{c|c|c}
\hline
\textbf{Parameter} & \textbf{LOS} & \textbf{NLOS} \\
\hline
Frame Repetition Factor  ($N_{f}$) & 20 & 20 \\
\hline
Doppler frequency ($f_{\text{Doppler}}$) & $150\,\mathrm{Hz}$ & $100\,\mathrm{Hz}$ \\
\hline
User speed ($v_d$) & $9\,\mathrm{m/s}$ & $6\,\mathrm{m/s}$ \\
\hline
Coherence time ($t_c$) & $2.82\,\mathrm{ms}$ & $4.23\,\mathrm{ms}$ \\
\hline
Block duration ($T_b$) & $32\,\mu\mathrm{s}$ & $32\,\mu\mathrm{s}$ \\
\hline
Inter-block correlation ($\rho$) & $0.9888$ & $0.9925$ \\
\hline
\end{tabular}
\label{tab:channel_params}
\end{table}
\vspace{-0.4cm}

%% file: sections/conclusion.tex
\section{Conclusion}
In this study, we proposed a novel iterative channel estimation, detection and decoding scheme, denoted ICCED, along with the ICCE channel estimation and the ICT channel tracking algorithms for RIS-assisted MIMO systems. Unlike existing approaches, ICCED exploits coding in the uplink to use parity bits for both decoding and channel estimation while employing encoded pilots to enhance performance. The proposed ICCED scheme, ICCE and ICT algorithms significantly reduce the minimum number of pilots required for both LOS and NLOS scenarios, achieving superior performance in LOS scenarios. Numerical results show that proposed approaches have large performance gains over competing techniques in terms of channel estimation NMSE and BER.

%% file: sections/appendix.tex
\appendices
\section{Detailed Analysis of MSE$_1$}
\label{app:lambda}
From~\eqref{eq:mse_1}, we express $\text{MSE}_1$ as
\begin{align}
\label{eq:appendix0}
\text{MSE}_1 &= M \sigma_z^2 \cdot \text{Tr} \Big\{ 
\underbrace{\mathbb{E} \left[ \left\| \boldsymbol{\Lambda} \boldsymbol{\Lambda}_{\text{ref}}^\dagger \right\|^2 \right]}_{\text{S}_1} \nonumber \\
&\quad - 2 \underbrace{ \mathbb{E} \left[ \boldsymbol{\Lambda} \boldsymbol{\Lambda}_{\text{ref}}^\dagger \right]}_{\text{S}_2} 
+ \mathbf{I}_{L_eK} 
\Big\}.
\end{align}

Using~\eqref{eq:lamblambH}, the term S$_2$ can be computed as
\begin{align}
\label{eq:appendix3}
\mathbb{E} \left[ \boldsymbol{\Lambda} \boldsymbol{\Lambda}_{\text{ref}}^\dagger \right] 
&= \mathbb{E} \left[ \frac{1}{\sigma_x^2 T} \sum_{i=1}^{T} 
\left( \boldsymbol{\Omega}^{(i)}_\text{kron} \right)^H 
\boldsymbol{\lambda}^{(i)} (\boldsymbol{\lambda}^{(i)})^H \right] \nonumber \\
&= \frac{1}{\sigma_x^2 T} \sum_{i=1}^{T} 
\mathbb{E}\left[ \left( \boldsymbol{\Omega}^{(i)}_\text{kron} \right)^H \right] 
\mathbb{E}\left[ \boldsymbol{\lambda}^{(i)} (\boldsymbol{\lambda}^{(i)})^H \right],
\end{align}
where the last step assumes independence.

Since \(\boldsymbol{\Omega}^{(i)}_\text{kron}\) is diagonal, and applying~\eqref{eq:expectedValue}, we obtain the expression in~\eqref{eq:appendix1}. To evaluate  S$_1$ , as shown in~\eqref{eq:appendix2}, we distinguish between two cases: \( i = j \) and \( i \neq j \), and utilize the identity \( \boldsymbol{\Omega}^{(i)}_\text{kron} 
\left( \boldsymbol{\Omega}^{(i)}_\text{kron} \right)^H = \mathbf{I}_{KL_e} \).

Finally, substituting the expressions~\eqref{eq:appendix3},~\eqref{eq:appendix1}, and~\eqref{eq:appendix2} into~\eqref{eq:appendix0}, we obtain
\begin{align}
    \text{MSE}_1 &=ML_eK\sigma_z^2\left[{L_e(K-(1-2p)^2)}/{T} +4p^2\right]
    \nonumber \\
    &={\mathbb{E}\left[\|\mathbf{Z}_{\mathrm{all}}\|_F^2\right]}\left[{L_e(K-(1-2p)^2)}/{T} +4p^2\right] \nonumber
\end{align}

%% file: sections/appendix2.tex
\begin{figure*}[!ht]
\begin{align}
\label{eq:appendix1}
\operatorname{tr}\left(\mathbb{E} \left[ \boldsymbol{\Lambda} \boldsymbol{\Lambda}_{\text{ref}}^\dagger \right]\right) 
&= \frac{(1 - 2p)}{\sigma_x^2 T} 
\sum_{i=1}^{T} \operatorname{tr}\left(\mathbb{E} \left[ \boldsymbol{\lambda}^{(i)} (\boldsymbol{\lambda}^{(i)})^H \right]\right) = \frac{(1 - 2p)}{\sigma_x^2 T} 
\sum_{i=1}^{T} \operatorname{tr}\left(\mathbb{E} \left[ 
(\mathbf{x}^{(i)} \otimes \boldsymbol{\varphi}_e^{(i)}) 
(\mathbf{x}^{(i)} \otimes \boldsymbol{\varphi}_e^{(i)})^H 
\right]\right) \hspace{6em} \nonumber \\
&= \frac{(1 - 2p)}{\sigma_x^2 T} 
\sum_{i=1}^{T}\operatorname{tr}\left(\mathbb{E} \left[ 
\mathbf{x}^{(i)} (\mathbf{x}^{(i)})^H 
\right] \otimes 
 \left[ 
\boldsymbol{\varphi}_e^{(i)} (\boldsymbol{\varphi}_e^{(i)})^H 
\right]\right) =
\frac{(1 - 2p)}{T}\sum_{i=1}^{T}\operatorname{tr}\left(\mathbf{I}_K\right) \operatorname{tr}\left(
\left[ 
\boldsymbol{\varphi}_e^{(i)} (\boldsymbol{\varphi}_e^{(i)})^H 
\right]\right) \nonumber\\
&= (1-2p)KL_e.
\end{align}

\begin{align}
\label{eq:appendix2}
\operatorname{tr}\left(\mathbb{E} \left[\left\| \boldsymbol{\Lambda} \boldsymbol{\Lambda}_{\text{ref}}^\dagger \right\|^2\right]\right) &
= \operatorname{tr}\left(\mathbb{E} \left[
\left( \boldsymbol{\Lambda} \boldsymbol{\Lambda}_{\text{ref}}^\dagger \right)^H 
\left( \boldsymbol{\Lambda} \boldsymbol{\Lambda}_{\text{ref}}^\dagger \right)
\right]\right) = \frac{1}{{\sigma_x^4 T^2}} \operatorname{tr}\left({\mathbb{E} \left[
\left( \sum_{i=1}^{T} 
\boldsymbol{\lambda}^{(i)} (\boldsymbol{\lambda}^{(i)})^H \boldsymbol{\Omega}^{(i)}_\text{kron} 
\right)
\left( \sum_{j=1}^{T} 
\left( \boldsymbol{\Omega}^{(j)}_\text{kron} \right)^H 
\boldsymbol{\lambda}^{(j)} (\boldsymbol{\lambda}^{(j)})^H 
\right)
\right]}\right) \nonumber \\
 &\overset{\underset{\mathrm{i=j}}{}}{\rightarrow} \sum_{i=1}^T \frac{1}{\sigma^4_xT^2}\operatorname{tr}\left(\mathbb{E}\left[
\mathbf{x}^{(i)} (\mathbf{x}^{(i)})^H\mathbf{x}^{(i)} (\mathbf{x}^{(i)})^H
\right] \otimes \left[\boldsymbol{\varphi}_e^{(i)} (\boldsymbol{\varphi}_e^{(i)})^H\boldsymbol{\varphi}_e^{(i)} (\boldsymbol{\varphi}_e^{(i)})^H \right]\right) \nonumber \\
& \overset{\underset{\mathrm{i=j}}{}}{\rightarrow}  \frac{1}{\sigma^4_xT^2}\sum_{i=1}^T\operatorname{tr}\left(\left[\sigma_x^4(K)\mathbf{I}_{K}\right]\right)\operatorname{tr}\left(\left[\boldsymbol{\varphi}_e^{(i)} (\boldsymbol{\varphi}_e^{(i)})^H\boldsymbol{\varphi}_e^{(i)} (\boldsymbol{\varphi}_e^{(i)})^H \right]\right) = \frac{(KL_e)^2}{T} \nonumber \\
\text{and,} \nonumber \\
&\overset{\underset{\mathrm{i\neq j}}{}}{\rightarrow}
\frac{(1-2p)^2}{\sigma^4_xT^2}
\sum_{\substack{i,j=1 \\ i \neq j}}^{T}\operatorname{tr}\left(\mathbb{E} \left[ 
(\mathbf{x}^{(i)} \otimes \boldsymbol{\varphi}_e^{(i)}) 
(\mathbf{x}^{(i)} \otimes \boldsymbol{\varphi}_e^{(i)})^H 
\right]
\mathbb{E} \left[ 
(\mathbf{x}^{(j)} \otimes \boldsymbol{\varphi}_e^{(j)}) 
(\mathbf{x}^{(j)} \otimes \boldsymbol{\varphi}_e^{(j)})^H 
\right]\right) \nonumber \\
&\overset{\underset{\mathrm{i\neq j}}{}}{\rightarrow}
\frac{(1-2p)^2}{T^2}\sum_{\substack{i,j=1 \\ i \neq j}}^{T}
\operatorname{tr}(\mathbf{I}_{K}) \operatorname{tr}\left(\left[\boldsymbol{\varphi}_e^{(i)} (\boldsymbol{\varphi}_e^{(i)})^H\boldsymbol{\varphi}_e^{(j)} (\boldsymbol{\varphi}_e^{(j)})^H \right] \right) = {(1-2p)^2}L_eK\left(1-\frac{L_e}{T}\right)
\end{align}
\hrulefill
\end{figure*}

%% file: sections/appendix_B.tex
\section{MSE Analysis of the Channel Refinement Error under LOS Conditions}
\label{app:B}

Using the results of [30], the channel estimation error can be modeled as
\begin{equation}
\widetilde{\mathbf{H}} = \mathbf{H} - \hat{\mathbf{H}},
\end{equation}
where each element is approximated as
\begin{equation}
\widetilde{h}_{i,j} \sim \mathcal{CN}\left(0, \sigma_h^2(1 - \varrho)\right),
\end{equation}
with
\begin{equation}
\varrho = \left(1 + \frac{\sigma_n^2}{2K \sigma_x^2 \sigma_h^2}\right)^{-1}.
\end{equation}
By substituting $\varrho$ into the previous expression, we obtain
\begin{equation}
\widetilde{h}_{i,j} \sim \mathcal{CN}\left(0, \frac{\sigma_h^2 \sigma_n^2}{\sigma_n^2 + 2K \sigma_x^2 \sigma_h^2}\right).
\end{equation}

Since
\begin{equation}
\mathbf{y} = \mathbf{H}\mathbf{x} + \mathbf{n} = \hat{\mathbf{H}}\mathbf{x} + \underbrace{\left(\widetilde{\mathbf{H}}\mathbf{x} + \mathbf{n}\right)}_{\mathbf{u}_e},
\end{equation}
the term $\mathbf{u}_e$ represents the combined effect of the channel estimation error and the additive noise.
We assume that $\widetilde{h}_{i,j}$ and the receiver noise $n \sim \mathcal{CN}(0, \sigma_n^2)$ are statistically independent, and that the transmitted vector $\mathbf{x}$ contains symbols drawn from a discrete uniform QPSK constellation. Consequently, the multiplication by $\mathbf{H}$ only rotates the symbol phases and does not alter their statistical distribution.
Hence, their sum can be expressed as
\begin{equation}
\mathbf{u}_e  \sim \mathcal{CN}\left(0, \left(\frac{K \sigma_h^2 \sigma_n^2}{\sigma_n^2 + 2K \sigma_x^2 \sigma_h^2} + \sigma_n^2\right)\mathbf{I}_M\right).
\end{equation}

To compute the MSE of the refined channel estimation, the noise matrix $\mathbf{U}$ in Eq.~(47) is replaced by $\mathbf{U}_e = [\mathbf{u}_e^{(i)}, \dots, \mathbf{u}_e^{(i+T-1)}]$.
This substitution does not affect the value of $\mathrm{MSE}_1$, while the second term becomes
\begin{equation}
\mathrm{MSE}_\text{2,LOS} = \frac{\sigma_n^2 \varrho_e}{(T - K L_e) \sigma_x^2 \sigma_z^2},
\end{equation}
where
\begin{equation}
\varrho_e = \left(\frac{K\sigma_h^2}{\sigma_n^2 + 2K \sigma_x^2 \sigma_h^2} + 1\right)
\end{equation}
is the contribution of the direct-link channel estimation error.

Finally, we compute the NMSE for the LOS case. 
In this scenario, let $e' \triangleq \mathrm{NMSE}_{\mathrm{LOS}}$. 
Assuming that the direct and cascaded channels are independent, 
it can be expressed as
\begin{equation}
\label{eq:nmse_eff}
e' = \frac{
    \mathbb{E}\!\left[
        \big\|
            \hat{\bar{\mathbf{H}}}_e
            - 
            \bar{\mathbf{H}}_e
        \big\|_F^2
    \right]
}{
    \mathbb{E}\!\left[
        \big\|
            \bar{\mathbf{H}}_e
        \big\|_F^2
    \right]
}
=
\frac{
    \mathrm{MSE}_{1}
    +
    \mathrm{MSE}_{2,\mathrm{LOS}}
}{
    \mathbb{E}\!\left[\|\mathbf{H}\|_F^2\right]
    +
    \mathbb{E}\!\left[\|\mathbf{Z}_{\mathrm{all}}\|_F^2\right]
}.
\end{equation}